\documentclass[reprint,aps,superscriptaddress,nofootinbib,groupedaddress,preprintnumbers]{revtex4-1}

\usepackage{amsmath,amsthm,amsfonts}
\usepackage{graphicx}
\usepackage{physics}
\usepackage{hyperref}
\hypersetup{
	colorlinks   = true, %Colours links instead of ugly boxes
	urlcolor     = blue, %Colour for external hyperlinks
	linkcolor    = blue, %Colour of internal links
	citecolor    = red   %Colour of citations
}

\newcommand{\nc}{\newcommand}

\nc{\calR}{{\cal{R}}}
\nc{\calP}{{\cal{P}}}
\nc{\cN}{ {\cal{N}} }

\nc{\Mpt}{M_{_{\rm Pl}}^2}

\usepackage{tikz}
\usetikzlibrary{arrows,shapes}
\usetikzlibrary{trees}
\usetikzlibrary{matrix,arrows} 				% For commutative diagram
\usetikzlibrary{positioning}				% For "above of=" commands
\usetikzlibrary{calc,through}				% For coordinates
\usetikzlibrary{decorations.pathreplacing}  % For curly braces
\usepackage{pgffor}							% For repeating patterns

\usetikzlibrary{decorations.pathmorphing}	% For Feynman Diagrams
\usetikzlibrary{decorations.markings}
\tikzset{
    vector/.style={decorate, decoration={snake}, draw},
	provector/.style={decorate, decoration={snake,amplitude=2.5pt}, draw},
	antivector/.style={decorate, decoration={snake,amplitude=-2.5pt}, draw},
    fermion/.style={draw=black, postaction={decorate},
        decoration={markings,mark=at position .55 with {\arrow[draw=black]{>}}}},
    fermionbar/.style={draw=black, postaction={decorate},
        decoration={markings,mark=at position .55 with {\arrow[draw=black]{<}}}},
    fermionnoarrow/.style={draw=black},
    gluon/.style={decorate, draw=black,
        decoration={coil,amplitude=4pt, segment length=5pt}},
    scalar/.style={dashed,draw=black, postaction={decorate},
        decoration={markings,mark=at position .55 with {\arrow[draw=black]{>}}}},
    scalarbar/.style={dashed,draw=black, postaction={decorate},
        decoration={markings,mark=at position .55 with {\arrow[draw=black]{<}}}},
    scalarnoarrow/.style={dashed,draw=black},
    electron/.style={draw=black, postaction={decorate},
        decoration={markings,mark=at position .55 with {\arrow[draw=black]{>}}}},
	bigvector/.style={decorate, decoration={snake,amplitude=4pt}, draw},
}

\tikzstyle{block} = [draw, rectangle, 
    minimum height=3em, minimum width=6em]

\begin{document}

\preprint{}

\title{ Alleviating $H_0$ and $S_8$ Tensions Simultaneously in K-essence Cosmology}

\author{Seyed Ali Hosseini Mansoori$^{1}$ and Hossein Moshafi$^{2}$}
\email{shosseini@shahroodut.ac.ir\\moshafi@ipm.ir}

\affiliation{$^{1}$Faculty of Physics, Shahrood University of Technology, P.O. Box 3619995161 Shahrood, Iran}

\affiliation{$^{2}$School of Astronomy, Institute for Research in Fundamental Sciences (IPM), Tehran, Iran, P.O. Box 19395-5531}

\begin{abstract}
The present work begins by examining the early-Universe inflationary epoch of a special K-essence model, which incorporates a linear coupling term between the scalar field potential and the canonical Lagrangian. For the power law potential, we both numerically and analytically prove that the inflationary parameters such as the spectral index and tensor-to-scalar ratio are compatible with the recent BICEP/Keck observations. Continuing this work, our analysis based on comparing early-Universe observations with late-Universe measurements indicates that the tension on the Hubble parameter $H_0$ and the growth of structure parameter $S_8$ can be alleviated simultaneously. More precisely, compared to the standard $\Lambda$CDM model, our model can reduce $H_0$ tension to roughly $2.2 \sigma$ and the $S_8$ discrepancy diminishes to $0.82\sigma$.

\end{abstract}

\maketitle %\pacs{98.80.Cq}

\section{Introduction}\label{intro}

The $\Lambda$CDM model has been successful in explaining various observations in cosmology and astrophysics across different scales and periods. These include the anisotropies in the Cosmic Microwave Background (CMB), the clustering of the large-scale structure (LSS), cosmic shear, the magnitude-redshift relation of distant Type Ia Supernovae (SNeIa), and light element abundances~\cite{SupernovaSearchTeam:1998fmf,SupernovaCosmologyProject:1998vns,DES:2017qwj,Planck:2018vyg,SPT:2019fqo,ACT:2020gnv,eBOSS:2020yzd,KiDS:2020suj,Mossa:2020gjc,Brout:2022vxf}. Nevertheless, it presents potential drawbacks, both theoretical and observational \cite{Perivolaropoulos:2021jda}. The first category includes issues like the non-renormalizability of general relativity and the cosmological constant problem. The second encompasses challenges such as the dynamic nature of dark energy, the onset of the inflationary phase, and a range of cosmological discrepancies. 

Among cosmological discrepancies, the "Hubble tension" is particularly noteworthy~\cite{Huterer:2023qez}. This mismatch is observed in several early- and late-time inferences of the Hubble constant $H_0$~\cite{Riess:2021jrx,Verde:2019ivm}. One of the most precise early-time inferences comes from CMB measurements by the \emph{Planck} satellite, which assuming $\Lambda$CDM, yields $H_0=(67.36 \pm 0.54)\,{\rm km}/{\rm s}/{\rm Mpc}$~\cite{Planck:2018vyg}. This value improves to $(67.62 \pm 0.47)\,{\rm km}/{\rm s}/{\rm Mpc}$ when combined with Baryon Acoustic Oscillation (BAO) and Hubble flow SNeIa data~\cite{Planck:2018vyg}. In contrast, one of the most precise local measurements comes from the SH0ES team, which yields $(73.04 \pm 1.04)\,{\rm km}/{\rm s}/{\rm Mpc}$ through a distance ladder making use of Cepheid-calibrated SNeIa~\cite{Riess:2021jrx}. The significance of the tension varies depending on the dataset considered but falls between $4\sigma$ and $6\sigma$ in most cases, which makes it one of the most exciting open problems in cosmology. %Various reviews and research papers have been published on this topic, providing insights and discussing possible explanations. (See~ for reviews).

Alongside the Hubble tension, there exists an additional discrepancy on the amplitude of matter fluctuations between early and late time cosmological probes, typically parameterized as $S_{8} \equiv \sigma_{8}\sqrt{\Omega_{m}/0.3}$, in which $\sigma_{8}$ is the root mean square of matter fluctuations on an $8 h^{-1}$Mpc  scale, and $\Omega_{m}$ is the total matter abundance. To put it in numbers, the value obtained from the CMB by \emph{Planck} \cite{Planck:2018vyg} $S_{8} = 0.834 \pm 0.016$ is $2.9 \sigma$ bigger than the directly measured values from KiDS-1000 cosmology \cite{Heymans:2020gsg} and DES Y3 \cite{DES:2021wwk} combined give $S_{8} = 0.769 \pm 0.016$. A plethora of models have been suggested to address one or both of these tensions (see for instance \cite{Riess:2019qba,DiValentino:2020zio,DiValentino:2021izs,Schoneberg:2021qvd,Shah:2021onj,Abdalla:2022yfr,DiValentino:2022fjm,Hu:2023jqc,Heisenberg:2022lob,Heisenberg:2022gqk,Braglia:2020iik,Lee:2022cyh,Krishnan:2020obg,Bernal:2016gxb,Vagnozzi:2019ezj,Guo:2018ans,Mortsell:2018mfj,Mortsell:2021nzg,Efstathiou:2020wxn,Braglia:2020iik,Petronikolou:2021shp,Basilakos:2023kvk,Gangopadhyay:2023nli,Sharma:2022oxh,Moshafi:2022mva,Poulin:2018cxd,Elizalde:2020mfs,Mostaghel:2018pia,Mostaghel:2016lcd,DeFelice:2020cpt,DiValentino:2024wgi,Abdalla:2022yfr,DiValentino:2020vvd,Benisty:2020kdt,Adi:2020qqf,Kazantzidis:2019nuh,Nunes:2018xbm,DAgostino:2023cgx,Braglia:2020auw,Bouche:2023xjw,Escamilla-Rivera:2022mkc,Petronikolou:2023cwu,Banihashemi:2018has,Ivanov:2020mfr,Keeley:2022ojz,Bose:2020cjb,Huterer:2023ldv,Vagnozzi:2023nrq,Petronikolou:2021shp,Petronikolou:2023cwu,yarahmadi2024using,Pedreira:2023qqt,Basilakos:2023kvk,Heisenberg:2022lob,Heisenberg:2022gqk,Vagnozzi:2023nrq,Kable:2023bsg,Frion:2023xwq,Tiwari:2023jle,Jedamzik:2020krr,Dainotti:2021pqg,Dainotti:2022bzg}).

Furthermore, in the realm of inflationary cosmology, recent findings from \emph{Planck} significantly limit the range of acceptable inflationary parameters at early time. For example, observations from \emph{Planck} and BICEP/Keck throughout the 2018 observation period \cite{BICEP:2021xfz} indicate that the tensor-to-scalar ratio parameter is constrained to $ r < 0.036 $ with $95 \%$ confidence. Due to this constraint, the standard chaotic inflation models \cite{linde1983chaotic} with a potential of $ \phi^n $, even for $n = 2/3$, have been ruled out \cite{BICEP:2021xfz}.   

The primary objective of this paper is to establish an inflation model in which the parameters $n_s$ and $r$ align with the latest observational constraints. For this purpose, we present a novel subset of K-essence models that includes the coupling between the potential function and the canonical Lagrangian~\cite{Armendariz-Picon:1999hyi,Armendariz-Picon:2000ulo}. This coupling term can influence the slow-roll parameters, thereby potentially altering the values of the spectral index $n_s$ and the tensor-to-scalar ratio $r$ at the CMB scale. We anticipate that our model will produce values for $n_s$ and $r$ that are consistent with  the current BICEP/Keck bound, rather than being excluded in the standard model of chaotic inflation.
The second aim of this study is to reconcile the discrepancies related to the Hubble constant $H_0$ and the structure growth rate $S_8$ by synthesizing diverse observational datasets. 

The paper is organized as follows: In Sec~\ref{sec:model} we first present our model setup. Subsequently, we explore inflationary solutions within our model and demonstrate that the values of the spectral index $n_{s}$ and the tensor-to-scalar ratio $r$ align with the latest BICEP/Keck constraints. In the rest of the sections, we attempt to show how our model can lead to the alleviation of both $H_0$ and $S_{8}$ tensions simultaneously. In order to accomplish this, we first introduce the observational data sets and the statistical methods used in this work in Sec~\ref{sec:data}. Section~\ref{sec:results} concentrates on presenting our results for this model in light of the combination of different datasets. Finally, we discuss findings in Sec~\ref{sec:conclusion}.

\section{Model}\label{sec:model}
Allow us to consider a special class of the \textit{K-essence} models \cite{Armendariz-Picon:2000ulo} which is introduced by
\begin{equation}\label{action}
P(X,\phi)= f(\phi)\mathcal{L}_{\rm cano}
\end{equation}
where $\mathcal{L}_{\rm cano}=X-V(\phi)$ is the canonical scalar field Lagrangian in which $X=-\partial_{\mu}\phi \partial^{\mu}\phi/2$ and $V(\phi)$ is the potential function. We assume that $f$ depends on the potential $V$ via
\begin{equation}
f(\phi)=1+(2 \mathcal{K}/M_{\rm pl}^4)V(\phi)
\end{equation}
 in which $M_{\rm pl}$ is the reduced Planck mass and $k$ is a dimensionless constant. 
 %Without loss of generality, one can take $k<0$.
 %\footnote{At the level of cosmological perturbations, one must consider $k<0$ and the upper bound on the potential as $V< 1/2|k|$ to remedy the ghost and gradient instabilities \cite{HosseiniMansoori:2023zop}.}%
  Note that as $\mathcal{K}=0$ the above function reduces to the canonical scalar field Lagrangian. When comparing it to the standard canonical k-essence model, where the kinetic and potential terms are sum-separable as $P(X,\phi)=X-V(\phi)$, there is an additional coupling term, such as $V(\phi) X$. This term may change the inflationary parameters results reported in the sum-separable canonical k-essence model. In the next section, we will therefore investigate the impact of such a term on inflationary parameters, such as the spectral index $n_{s}$ and the tensor-to-scalar ratio $r$, through both analytical and numerical methods. 

\section{Inflationary parameters }\label{subsec:chaotic-slowroll}
For the k-essence function \eqref{action}, the background equations of motion in a spatially flat FLRW spacetime are obtained by
 \begin{eqnarray}
  && 3 H^2=f (X+V) \label{fir1}\\
 \dot{X} f&+&\dot{\phi} \Big(f_{,\phi}(X+V)+fV_{,\phi}\Big) -6 f H X \label{fir2}
 \end{eqnarray} 
 where the prime stands for the derivative with respect to cosmic time $t$, $H=\dot{a}/a$ is the Hubble parameter, and $X=\dot{\phi}^2/2$. Moreover, we set $M_{\rm pl}^2=1$ throughout this paper for convenience.
 
As we consider the slow-roll approximation, where $\dot{\phi} \ll V$ (or $X \ll V$) and $\ddot{\phi} \ll H \dot{\phi}$ (or $\dot{X} \ll H X$), Eqs. \eqref{fir1} and \eqref{fir2} reduce to\footnote{Note that, in the slow-roll limit, the $\mathbb{T}^2$-inflation model introduced in Refs. \cite{HosseiniMansoori:2023zop,HosseiniMansoori:2023mqh} is equivalent to our present model.} 
\begin{eqnarray}\label{Fridmannnew1}
3H^{2}&\simeq & f\,\ V\\
\dot{\phi}V_{,\phi}\left( 2 f-1 \right) & \simeq & -6 X f H.\label{continuitynew1}
\end{eqnarray}
From Eq. \eqref{Fridmannnew1}, one finds there is an upper bound on the potential, i.e., $f>0$. This condition ensures that  the model \eqref{action} is free from ghost and gradient instabilities at the level of cosmological scalar perturbations \cite{Chen:2006nt}. Without loss of generality, we can assume $\mathcal{K}<0$. As a result, form the condition $f>0$, we obtain an upper bound on the potential function, given by $V<1/2| \mathcal{K}|$.

As long as one differentiates both sides of Eq. (\ref{Fridmannnew1}) with respect to time and combines it with Eq. (\ref{Fridmannnew1}), the Hubble slow-roll parameter reads as
\begin{equation}\label{epsilonv1}
\varepsilon_{H} \equiv -\frac{\dot{H}}{H^2}\simeq -\frac{1}{2 }\Big(\frac{V_{,\phi}}{V }\Big)\Big(\frac{\dot{\phi}}{H}\Big)\Big(\frac{2f-1 }{f}\Big).
\end{equation}
 On the other hand, using Eq. \eqref{continuitynew1}, one can find that
\begin{equation}\label{phidoth}
\frac{\dot{\phi}}{H} = -\Big(\frac{V_{,\phi}}{V }\Big)\Bigg[\frac{2f-1 }{f^2}\Bigg]
\end{equation}
By combing the above relations, the slow roll parameter \eqref{epsilonv1} is rewritten as a function of the potential and its derivative, namely
\begin{equation}\label{epsilonvnew}
\varepsilon_{H} = \frac{1}{2 }\bigg(\frac{V_{,\phi}}{V }\bigg)^{2}\Bigg[\frac{\Big(2f-1 \Big)^{2}}{f^{3}}\Bigg].
\end{equation}
Analogous to the standard canonical model, in our model, the sound speed $c_{s}$  equals the speed of light, i.e., $c_{s}=1$ \cite{Chen:2006nt}.
%\begin{equation}\label{speed1}
% c_{s}^2=\frac{P_{,X}}{P_{,X}+2X P_{,XX}}= 1
 %\end{equation} 
%Clearly, the sound speed value is less than the speed of light and it may enhance the non-gaussianity in the model. 
%Note that all mentioned above parameters reduce to the standard form \cite{Li:2012vta} as $k \to 0$.
Furthermore, during the inflation era, it is necessary for the slow-roll parameters, namely $\varepsilon_{H}$ and $\eta_{H}\equiv \dot{\varepsilon_{H}}/(H \varepsilon_{H})$, to be much smaller than one. This condition holds for at least 50-60 e-folds to address the flatness and horizon problems. It is worth noting that the inflation terminates when either of the slow-roll parameters approaches unity.

%By making use of Eq. \eqref{phidoth}, the above relations can be expressed as a function of the potential and its derivatives.

Taking advantage of Eq. \eqref{phidoth}, we can obtain the number of e-foldings as 
\begin{equation}
N=-\int_{t_{e}}^{t} H dt=\int_{V_{e}}^{V} \frac{1}{2  \varepsilon_{H}} \frac{dV}{V} \Big(\frac{2f-1}{f}\Big)
\end{equation}
 in which the subscript ``$e$'' stands for the value of the quantities at the end of the inflation. Let us now choose a simple potential function like the chaotic potential with $V(\phi)=(A/M_{\rm pl}^{n}) \phi^{n}$, in which $A$ is a dimensionless coefficient and $n$ is a rational number. It's important to note that $A$ represents the normalization parameter, which is determined by the amplitude of the scalar power spectrum at the CMB pivot scale ($k_{\rm CMB}=0.05 \rm Mpc^{-1}$), i.e. $\mathcal{P}_{CMB} \sim 2.1 \times 10^{-9}$. Therefore, the above expression converts to
\begin{equation}\label{Nfunction}
N=\frac{1}{2 n}\Big(\frac{V}{A}\Big)^{\frac{2}{ n}}\Big[1+\frac{ 2\mathcal{K} V}{2+ n} \Big(1- {}_{2}F_{1}(1,1+\frac{2}{ n},2+\frac{2}{ n},-4 \mathcal{K}  V)\Big)  \Big] 
\end{equation}
where ${}_{2}F_{1}$ is the Gaussian hypergeometric function. Moreover, the potential $V$ has been calculated at the beginning of the inflation. Because of the complex nature of the hypergeometric function, it is not possible to figure out the function $V$ in reverse as a function of $N$. However, achieving this can be possible by selecting smaller values of the potential, ensuring that $ |\mathcal{K}| V \leq \mathcal{O}(\sqrt{\varepsilon_{H}^{\rm cano}})<1/2$.  In this regime, using \eqref{Nfunction} one can derive 
 \begin{eqnarray}
 \label{VAexpand}
\frac{V}{A}\simeq \Big(2n N\Big)^{\frac{ n}{2}}\Big[1-\frac{1}{4}\Big(\frac{(2n)^{n+1} N^{n}}{(1+n)}\Big)\epsilon^2+\mathcal{O}(\epsilon^3)\Big] 
 \end{eqnarray}
where $\epsilon=\mathcal{K} A$. Although we have examined the aforementioned relationship up to the second order, as $\mathcal{K} V$ approaches the limit defined by $\sqrt{\varepsilon_{H}^{\rm cano}} \sim \mathcal{O}(0.01)$ \cite{maldacena2003non}, it becomes imperative to consider higher orders, such as the fourth order, to achieve a strong convergence between analytical and numerical results.

By taking into account Eq. \eqref{epsilonvnew} and Eq. \eqref{VAexpand} together, one writes down the
 slow-roll parameters and the sound speed as a function of $N$, namely 
 \begin{eqnarray}\label{slowrollN}
\nonumber \varepsilon_{H} &\simeq&  \frac{n}{4 N}\Big[1+(2n N)^{\frac{n}{2}} \epsilon-\frac{ 4(2 n N)^{n} (1+2n)}{1+n} \epsilon^2
 +\mathcal{O}(\epsilon^3)\Big]\\
\nonumber \eta_{H}&\simeq & \frac{1}{ N}\Big[1-n (2 n N)^{\frac{n}{2}} \epsilon+ \frac{2 n (2 n N)^{n}(3+5n)}{1+n} \epsilon^2
 +\mathcal{O}(\epsilon^3)\Big]
 \end{eqnarray} 
Regarding all the relations mentioned above, the formal solution for the spectral index \eqref{nsanalatic} is obtained to be\footnote{In K-essence models, the scalar and tensor power spectra within the slow roll regime are defined as \cite{chen2007observational,Seery:2005wm}:
\begin{equation}\label{slopower}
\mathcal{P}_{\mathcal{R}} \simeq \frac{1}{8 \pi^2} \frac{H^2}{\varepsilon_{H} }|_{k=aH}, \hspace{1cm}\mathcal{P}_{h}=\frac{2}{\pi^2} H^2|_{k=aH}
\end{equation}

Afterward, one can compute the spectral index, $n_s$, and the tensor-to-scalar ratio, $r$ as 
\begin{eqnarray}\label{nsanalatic}
n_{s}-1&\equiv & \frac{d \ln \mathcal{P}_{\mathcal{R}}}{d \ln k}\simeq -2 \varepsilon_{H}-\eta_{H}\\
r &\equiv & \frac{\mathcal{P}_{h}}{\mathcal{P}_{\mathcal{R}}}=16 \varepsilon_{H} \label{rformula}
\end{eqnarray} } 
 \begin{eqnarray}\label{nsanaltic}
&&  n_{s}-1 \simeq  -\frac{1}{N}\Big[\frac{2+n}{2}-\frac{n(2nN)^{\frac{n}{2}} (n-2)}{6 N} \epsilon\\
\nonumber &+& \frac{n (2 n N)^{n}}{18(1+n)N^2} \Big(n(1+n)(n-2)+36 (2+3n)N^2\Big) \epsilon^2
 \\
\nonumber  &+&  \mathcal{O}(\epsilon^3)\Big]
 \end{eqnarray}
When comparing with Refs. \cite{Li:2012vta, Unnikrishnan:2012zu}, the spectral index undergoes modifications of the orders of $\epsilon$. Additionally, from Eqs. \eqref{slowrollN} and \eqref{rformula}, it becomes trivial to derive $r$ as a function of $N$.

Over the past few years, numerous observational constraints have been derived for the quantities $r$ and $n_{s}$ from various data sources, such as the \emph{Planck} 2018 data \cite{Planck:2018jri,BICEP2:2018kqh}, as well as the BICEP/\textit{Keck} (BK15~\cite{aghanim2020planck} and BK18 \cite{ade2021improved}). These constraints have imposed substantial limitations on the model's free parameters. For example, to be consistent with the current BICEP/Keck constraints \cite{ade2021improved}, in the $V=A \phi^{2/3}$ case, the value of the $\epsilon$ parameter should lie within the following range for $N=60$.
\begin{align} \label{eq:epsilon-bounds}
 -0.06903 < \epsilon < -0.0071
 \end{align}

\begin{figure}[ht]
\includegraphics[trim={1mm 1mm 0 0},clip,scale=0.25]{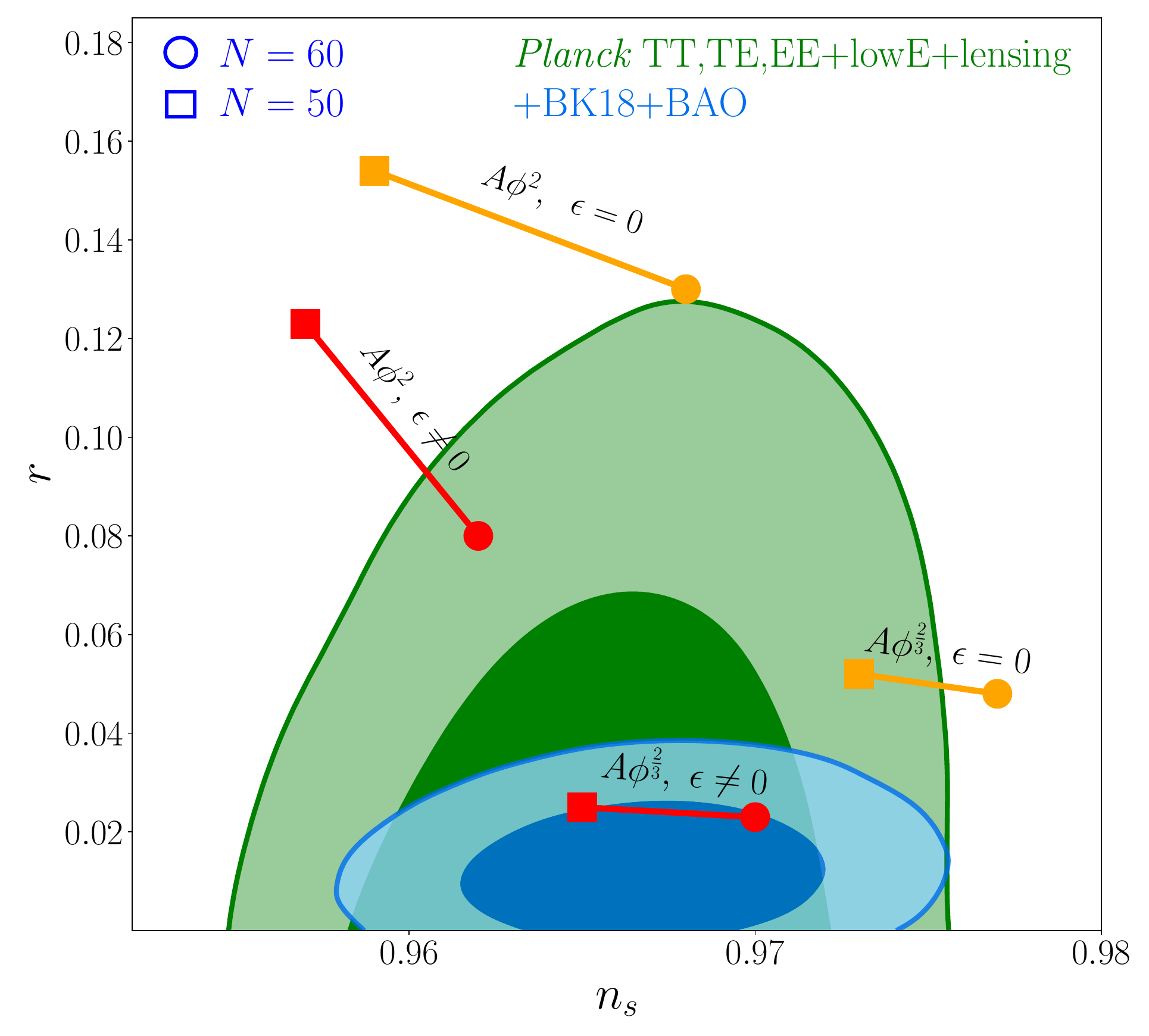}
 \caption{Marginalized joint $68\%$ and $95\%$ CL regions for $n_s$ and $r$ for the $\phi^{2/3}$ and $\phi^2$ models from \emph{Planck} 2018+lensing in comparison with BK18+BAO data~\cite{ade2021improved}}
 \label{fig:r-ns}
\end{figure}

As illustrated in Fig. \ref{fig:r-ns},  the predicted values of $\{r,n_{s}\}$ in the $\phi^{2/3}$ model for $\epsilon=-0.086$ fall entirely within the region determined by the BK18 results~\cite{ade2021improved},  as opposed to being ruled out in the standard model of chaotic inflation with $\epsilon=0$. Moreover, the numerical values of $\{n_{s},r\}$ on CMB scales, $\{0.971713,0.0176352\}$, closely match the analytical result given by Eq. \eqref{nsanaltic} for $N=60$.
 Obviously, the $\phi^2$ model still remains inconsistent with current observational data. Consequently, in the remainder of  this manuscript, we focus on the potential $V = A \phi^{2/3}$.

By analyzing the primordial power spectrum of scalar perturbations, we can understand how the $\epsilon$ parameter affects early-Universe observational predictions for the Hubble parameter $H_{0}$ and the growth of structure parameter $S_8$. 
The primordial power spectrum is expressed as
\begin{align} \label{eq:pps}
\mathcal{P}_s(k) = A_s \left(\frac{k}{k_{\rm CMB}}\right)^{n_s -1}
\end{align}
where $A_s$ represents the amplitude of scalar perturbations and the spectral index $n_{s}$ is derived Eq. \eqref{nsanaltic}, expressed as a function of the number of e-folds. Notice that $n_{s}$ is written as a function of the wave number $k$ during the inflation era where the Hubble parameter is roughly constant. 

%\textcolor{orange}{If we want to see how the $\epsilon$ parameter affects CMB predictions, we need to consider a compatible range with \emph{Planck} observations. Our proposed range for this parameter is as follows:
%\begin{align}
%-0.01 <\epsilon < -0.0001
%\end{align}
%Moreover, our model includes tensor perturbations, and we let the tensor-to-scalar ratio, $r$ parameter be a free parameter in the following range:
%\begin{align}
%0.001 < r < 0.2
%\end{align}}

%\textcolor{orange}{By analyzing the primordial power spectrum of scalar perturbations, we can understand how the $\epsilon$ parameter affects observational predictions. With a compatible range with \emph{Planck} observations, we can see how the $\epsilon$ parameter affects CMB predictions. Our model includes tensor perturbations, and we let the tensor-to-scalar ratio, $r$ parameter be a free parameter in the range of 0.001 to 0.2. With these considerations, we can effectively constrain the model with observational data and make accurate predictions about the early universe.}

Given the early power spectrum \eqref{eq:pps} and the $r$ value, we attempt to reduce $H_{0}$ and $S_{8}$ tensions between early-Universe results for our model and local late-Universe measurements in the upcoming sections. 

%For example, the numerical values of $\{n_{s},r\}$ on CMB scales are $\{0.971802,0.0176263\}$ which are in close agreement with analytic result \eqref{nsanaltic} as one takes $kA=-0.086$ and $N=60$.  

%%%%%%%%%%%%%%%%%%%%%%%%%%%%

 %%%%%%%%%%%%%%%%%%%%%%%

\begin{figure}
        \includegraphics[width=\linewidth]{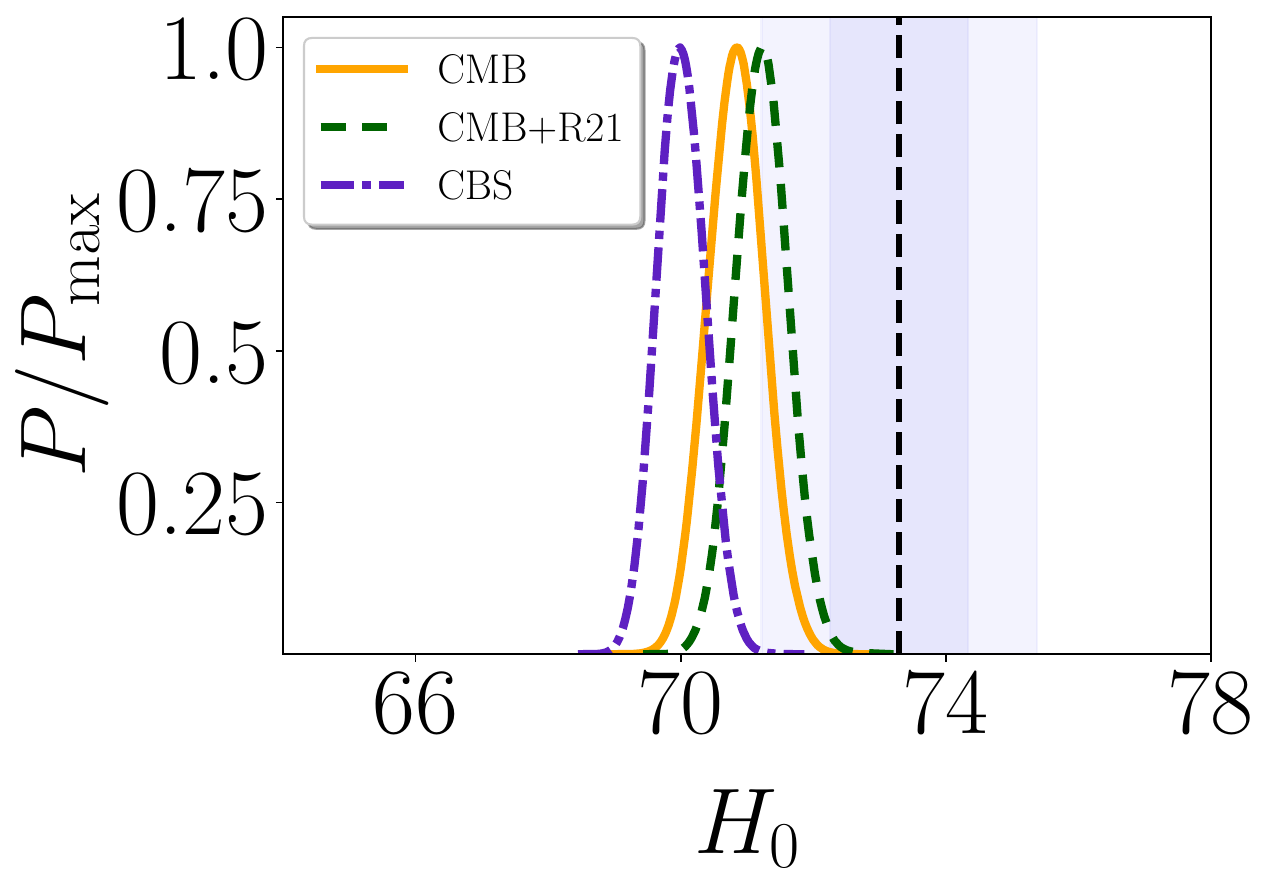}
        
        \includegraphics[width=\linewidth]{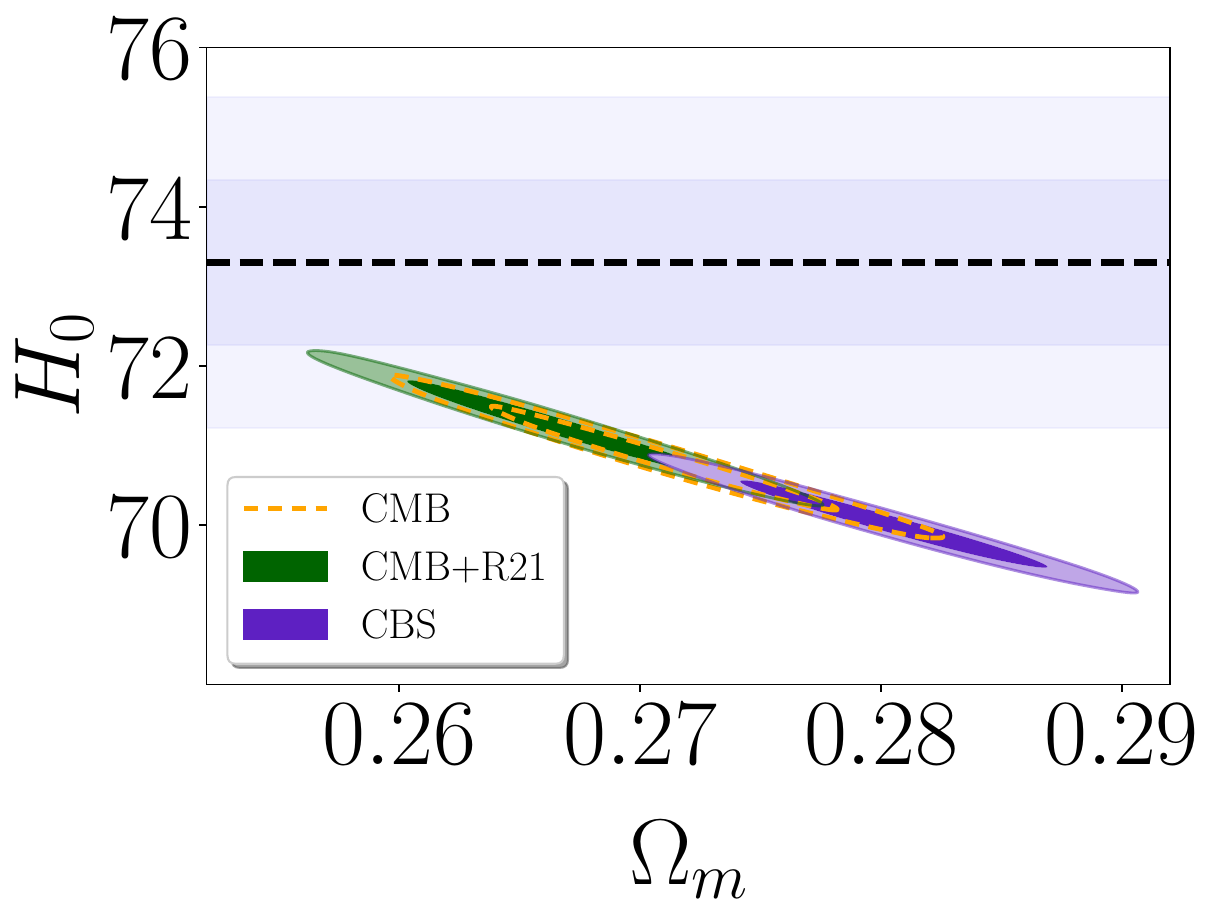}
    
    \caption{\footnotesize Alleviating the Hubble Tension in $\phi^{2/3}$ model. \textit{(Upper:)} one-dimensional posteriors of $H_0$ parameter are shown for the different combination of dataset. \textit{(Lower:)} $68\%$ and $95\%$ confidence level constraints on $H_0$ vs. $\Omega_m$ for the model. The shaded area shows the measurement of $H_0$ done by the SH0ES team and its $1\sigma$ and $2\sigma$ errors~\cite{Riess:2021jrx}.}
    \label{fig:H0-omegam}
\end{figure}
 \section{Alleviating $H_0$ and $S_{8}$ tensions}
\subsection{Methodology and Data}
\label{sec:data}

Before we proceed further, let us briefly describe the observational analysis tools and likelihoods used in this study. Using a modified version of the publicly available Code for Anisotropies in the Microwave Background, \texttt{CAMB}~\cite{Lewis:1999bs,Howlett2012} we do a Maximum Likelihood Analysis (MLA) employing the Cosmological Markov Chain Monte Carlo (MCMC) code \texttt{CosmoMC}~\cite{Lewis:2002ah}.
For the $\Lambda$CDM+r model, we have used the following 7 parameter space:

\begin{eqnarray}
\mathcal{P}_0 \equiv\Bigl\{ \Omega_{b}h^2, \Omega_{c}h^2, 100\theta_{\rm MC}, \tau, n_s, \ln[10^{10}A_{s}], r \Bigr\}~,
\label{eq:lcdm-parameter-space}
\end{eqnarray}
For our model, we have the same number of parameters:
\begin{eqnarray}
\mathcal{P}_1 \equiv\Bigl\{\Omega_{b}h^2, \Omega_{c}h^2, 100\theta_{\rm MC}, \tau, \epsilon, \ln[10^{10}A_{s}], r \Bigr\}~,
\label{eq:parameter-space}
\end{eqnarray}
where $\epsilon$ and $A_s$ represent the model parameters, $\tau_{\rm re}$ is the reionization optical depth, $n_s$ is the scalar spectral index, $A_{s}$ is the amplitude of the scalar primordial power spectrum, and $\Theta_{\rm MC}$ is an approximation of $\theta_*$, which represents the angular scale of the sound horizon at decoupling. We always consider a flat Universe ($\Omega_{\rm K}=0$). The priors assumed for parameters are summarized in Table~\ref{tab:priors}.

Our reference datasets in the study of $\phi^{2/3}$ model are the following:
\begin{enumerate}
\item {\bf CMB}: We use the most precise full-sky measurements of Cosmic Microwave Background (CMB) radiation performed by \emph{Planck} satellite. We use both high-$\ell$ temperature and polarization angular power spectra from the release of  ``\emph{Planck 2018}'' baseline \texttt{PLIK-TTTEEE} along with \emph{Planck} \texttt{low}-$\ell$ and \texttt{low}-E \texttt{(SimAll)} ($\ell \leq 30$)~\cite{Planck:2018vyg, Planck:2018lbu, Planck:2019nip}. We mention all of \emph{Planck} data (including temperature and polarization) by ``CMB''.

\item{\bf R21}: To assess whether our model can resolve the $H_0$ tension, we have included a Gaussian prior in the form of $H_0=73.30 \pm 1.04$ $\rm{km}/ \rm{s}~ \rm{Mpc}$, which was reported by the SH0ES team~\cite{Riess:2021jrx}. We refer to this prior as ``R21''.

\item {\bf BAO}: We also consider the various measurements of the Baryon Acoustic Oscillations (BAO) from different galaxy surveys as the \emph{Planck} collaboration in their 2018 analysis~\cite{Planck:2018vyg}, i.e. 6dFGS~\cite{Beutler:2011hx}, SDSS-MGS~\cite{Ross:2014qpa}, and BOSS DR12~\cite{BOSS:2016wmc}. We mention all these data points by ``BAO''. 

\item {\bf SN}: We utilize luminosity distance measurements of 1,048 Type Ia supernovae in the redshift range $z \in [0.01, 2.3]$ from the Pantheon sample~\cite{Pan-STARRS1:2017jku}. We refer to this catalog as ``SN''.

\item{ \bf DES}: We also use weak-lensing data from Year 3 of the Dark Energy Survey (DES-Y3) to reconcile the tension in $S_8$ parameter~\cite{DES:2021wwk}.

\end{enumerate}
In some analyses, we combine "CMB+BAO+SN" data and refer to it as ``CBS''.

To test the model for Hubble tension, we use the Gaussian Tension (GT), which applies the ``rule of thumb difference in mean''.~\cite{Raveri:2018wln,Schoneberg:2021qvd,Moshafi:2024guo}

%%%%%%%%%%%%%%%%%%%%%%%%%%%%%%%%%%%%%%%%%%%%%%%%%%%%
\begin{table*}[t]
	\centering
	\begin{tabular} {c || c c c c c c c }
 \hline
Parameters & {$\epsilon$} & {$\ln \left[10^{10} A_s \right]$} &{$r$} & {$\Omega_b h^2$} & {$\Omega_c h^2$} & {$\tau_{\rm re}$} & {$100 \Theta_{\rm MC}$} \\
%		Parameter & Priors      \\
		\hline
		\hline
	Priors &	$[-0.01, -0.0001]$  & $[3.7, 5.0]$  & $[0.001, 0.2]$ & $[0.005, 0.1]$ & $[0.001, 0.99]$ & $[0.01, 0.8]$ & $[0.5, 10]$\\

		\hline
	\end{tabular}
	\caption{Flat priors used on various free parameters of $\phi^{2/3}$ model, during statistical analysis. The $\epsilon$ and $r$ parameters are chosen to be compatible with the range in Eq.~\ref{eq:epsilon-bounds} and the recent BICEP/\emph{Keck} observation~Fig.~\ref{fig:r-ns}.}
	
	\label{tab:priors}
\end{table*}

%%%%%%%%%%%%%%%%%%%%%%%%%%%%%%%%%%%%%%%%%%%%%%%%
%%%%%%%%%%%%%%%%%%%%%%%%%%%%%%%%%%%%%%%%%%%%%%%%

\begin{table*}[tb]
\begin{center}
\resizebox{\textwidth}{!}{
\begin{tabular}{ c |c c c ||c c c }
\hline
\hline
 & $\phi^{2/3}$ model & $\phi^{2/3}$ model & $\phi^{2/3}$ model & $\Lambda$CDM & $\Lambda$CDM & $\Lambda$CDM \\ \hline
Parameters & CMB & CMB+R21 & CMB+BAO+SN & CMB & CMB+R21 & CMB+BAO+SN \\ \hline

{\boldmath$\epsilon$} & $< -0.00443$ & $< -0.00427$ & $< -0.00518$ & $0$ & $0$& $0$ \\

{\boldmath${\rm{ln}}(10^{10} A_s) $} & $4.388^{+0.019}_{-0.021}$ & $4.388^{+0.018}_{-0.023}$& $4.388^{+0.018}_{-0.023}$ & $3.044\pm 0.016$ & $3.047^{+0.015}_{-0.018}$& $3.044\pm 0.017$ \\

{\boldmath$\Omega_m $} & $0.2709\pm 0.0047$ & $0.2669\pm 0.0042$  & $0.2806\pm 0.0042$ &
 $0.3155\pm 0.0088$ & $0.2946\pm 0.0068$  & $ 0.3094^{+0.0053}_{-0.0061} $ \\

{\boldmath$H_0$}& $70.84\pm 0.42$&$71.21\pm 0.39$  & $70.00\pm 0.37$ & $67.34\pm 0.63$ & $68.93^{+0.50}_{-0.57}$  & $67.78\pm 0.43$ \\

$S_8$ & $0.758\pm 0.012$& $0.749\pm 0.012$& $0.778\pm 0.011$ & $0.832\pm 0.017 $ &$0.796\pm 0.014$ & $0.821\pm 0.013 $ \\

{\boldmath$r $} & $0.047\pm 0.020$ &$0.040^{+0.018}_{-0.023}$& $0.045^{+0.021}_{-0.033}$ & $ < 0.0549$ & $ < 0.0723$ & $< 0.0577 $ \\

		{\boldmath$\tau $} & $0.0699^{+0.0089}_{-0.010}$ & $0.0706^{+0.0086}_{-0.011}$ &$0.0684^{+0.0082}_{-0.011}$ & $ 0.0541\pm 0.0078$ & $ 0.0597^{+0.0075}_{-0.0089}$ & $0.0557\pm 0.0081$ \\
        \hline
		{\boldmath$\chi^2_{\rm R21}$} & $-$ & $4.2$ & $-$ & $-$ & $17.9$ & $-$\\
		\hline
%		{\boldmath$\chi^2_{\rm DES}$} & $-- $ & $--$ & $--$ &$--$ \\
%		\hline
		{\boldmath$\chi^2_{\rm BAO}$} & $-$ & $-$ & $13.4$ & $-$ & $-$ & $5.93 $ \\
		\hline
		{\boldmath$\chi^2_{\rm SN}$} & $-$ & $-$ & $1035.49$ & $-$ & $-$ & $1035.03$ \\
		\hline
		{\boldmath$\chi^2_{\rm CMB}$} & $2834.3$ & $2835.7$ & $2837.7$ & $2781.9 $ & $ 2788.3$ & $2781.9 $ \\

		\hline
		\hline

		{\boldmath$\chi^2_{\rm TOT}$} & $2834.3$ & $2839.9$ & $3886.59$ & $2781.9$ & $2806.2$ & $3822.86$ \\	
\hline
\hline

\end{tabular}
}
\end{center}
\caption{$68\%$ CL constraints for the $\phi^{2/3}$ model and the standard $\Lambda$CDM model, for CMB, CMB+R21 and CMB+BAO+SN. }
\label{tab:table_results}
\end{table*}

%%%%%%%%%%%%%%%%%%%%%%%%%%%%
\begin{figure}[ht]
\centering
 \includegraphics[width= \linewidth]{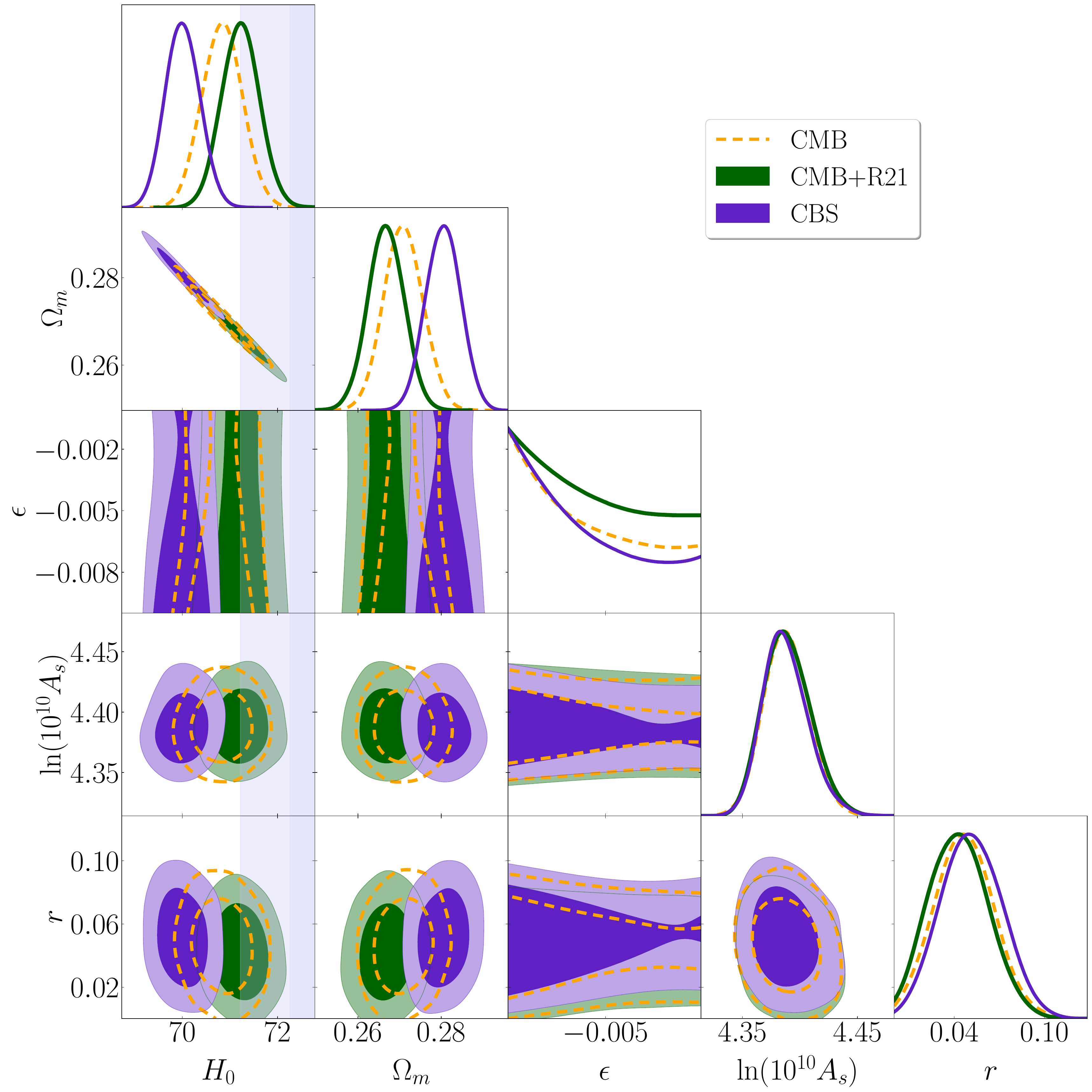}
 \caption{\footnotesize Marginalized joint $68\%$ and $95\%$ CL regions for the cosmological parameters in $\phi^{2/3}$ model with \emph{Planck}TT,EE, TE, with the EE likelihood at low multipoles joint with R21 prior on $H_0$ and BAO+SN data. The shaded area depicts the 1$\sigma$ and 2$\sigma$ error bars of the Hubble parameter measured by~\cite{Riess:2021jrx}.
 \label{fig:triangle-phi23-diff-data}}
\end{figure}
%%%%%%%%%%%%%%%%%%%%%%%%%%%%
\vspace{2cm}

\begin{figure}[ht]
 \includegraphics[width= .9\linewidth]{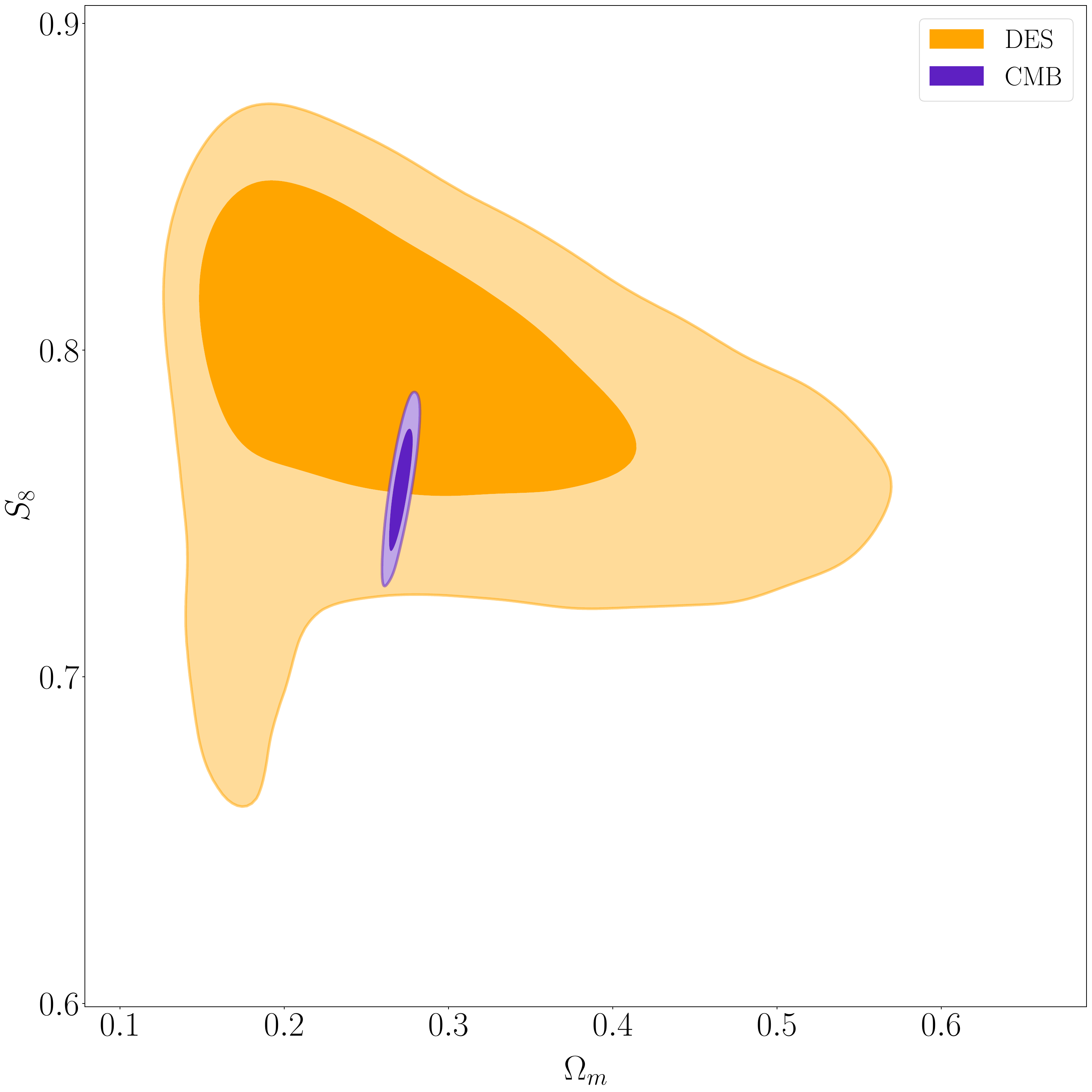}
 \caption{ \footnotesize 2D contour plots at $1\sigma$ and $2\sigma$ confidence levels for $S_8$ vs. $\Omega_m$ for the $\phi^{2/3}$ model to compare analysis based on CMB data and DES data.  \label{fig:s8-omegam}}
\end{figure}

\subsection{Results}
\label{sec:results}

To put our model to the test, we applied different combinations of observational data including Early-universe data (CMB), Standard Rulers (BAO), Standard Candles (SuperNovae), and also a prior of $H_0$ parameter extracted by SH0ES team~\cite{Riess:2021jrx}. Our analysis shows that our model can improve the discrepancy between $H_0$ values inferred from CMB observation indirectly and direct measurement of the Hubble rate from SuperNovae. We can see this result in Fig.~\ref{fig:H0-omegam}, which represents the prediction of our model for $H_0$ parameter is almost consistent with late-time measurement of Hubble parameter called ``R21"~\cite{Riess:2021jrx}. In this figure, the shaded area represents the measurement of the $H_0$ parameter according to the SH0ES team. The upper panel of Fig.~\ref{fig:H0-omegam} shows the comparison of the marginalized posterior distribution of $H_0$ parameter for the $\phi^{2/3}$ model from \emph{Planck 2018} temperature and polarization (CMB), \emph{Planck 2018}+R21 (CMB+R21) and \emph{Planck 2018}+BAO+SuperNovae (CBS). Also, in the lower panel of the same figure, we have plotted marginalized two-dimensional $1\sigma$ and $2\sigma$ confidence level regions for $H_0$ vs. $\Omega_m$ parameter. The two-dimensional contour plots of $H_0$ parameter vs. density parameter of matter $\Omega_m$ show that by applying different data combinations including CMB, CMB+SN, CMB+R21, and CMB+SN+R21 to our model, it shows consistency with ``R21" measurement completely. Our result shows that this model can relieve the tension in the Hubble parameter.

We summarize the results analysis of the $\phi^{2/3}$ model by comparing similar analysis for the standard model of cosmology $\Lambda$CDM in Table.~\ref{tab:table_results}. Remarkably, our analysis establishes an upper bound on the $\epsilon$ parameter across all scenarios involving CMB, CMB+R21, and CMB+BAO+SN (CSB) data. These upper bounds align with the limits specified in~Eq.~\ref{eq:epsilon-bounds}. In addition, by noting that both models have the same number of parameters, we see that $\phi^{2/3}$ model gives a higher value for $H_0$ parameter when we apply only CMB data. Also, this behavior is visible when we combine CMB data with SH0ES prior (R21) and also with BAO and SN data. Another notable result is the reduction in the $S_8$ parameter for the $\phi^{2/3}$ model when we apply only CMB data. This model in contrast to $\Lambda$CDM model puts a tighter constraint on the tensor-to-scalar ratio of primordial perturbations. By looking at the $\chi^2$ rows, we see that our model gives a higher $\chi^2$ value with respect to $\Lambda$CDM model. We can see a similar result in Fig.~\ref{fig:triangle-phi23-diff-data} which we depict marginalized joint $68\%$ and $95\%$ CL regions for cosmological parameters of the model with CMB, CMB+R21, and CMB+BAO+SN combination of data.

According to this table, by using the definition of Gaussian Tension~\cite{Moshafi:2024guo}, our model can reduce the Hubble parameter tension  to ~$2.2 \sigma$, namely
\begin{align}
H_0 &=70.84\pm 0.42 \quad \quad (68\% \rm{CL})&&
\end{align}
Our achievement was made possible by maintaining an identical number of free parameters as the standard model.

Moreover, the present model is capable of resolving the $S_8$ tension as we compare the analysis results based on either CMB data or only DES data. To assess a model's ability to alleviate the $S_8$ tension, it is necessary to compare the analysis results of the model with both CMB data and DES data. We cannot rely on the values obtained by considering only the $\Lambda$CDM model. Furthermore, the $\phi^{2/3}$ model has demonstrated that it can reduce the $S_8$ tension to the $0.82 \sigma$ level.

\begin{align}
S_8 &=0.758\pm 0.012\quad \quad(68\% \rm{CL})~~\emph{Planck}&&\\\nonumber
S_8 &=0.790\pm 0.037\quad \quad(68\% \rm{CL})~~\rm{DES}&&
\end{align}

This outcome reduces the tension in the $ S_8$ parameter, which is illustrated by the marginalized two-dimensional contours in Fig.~\ref{fig:s8-omegam}.

\section{Concluding Remarks}
\label{sec:conclusion}
In this study, we introduced a novel subset of the K-essence model. This variation contains a coupling term between the potential function and the kinetic term, represented as $V X$. The presence of this term has a significant impact on the inflationary parameters, leading to changes in our predictions for the scalar spectral index and the tensor-to-scalar ratio at CMB scales. These predictions differ from those of the standard chaotic inflation model.

We have investigated the impact of coupling on observations of the early and late Universe. Our model, similar to the standard cosmological model, has six parameters. However, we have included an extra parameter for the tensor-to-scalar ratio, $r$, in our analysis. We used different data combinations, including \emph{Planck 2018}, \emph{Planck}+R21, and \emph{Planck}+BAO+SN data to compare our results with those from the standard cosmological model. Our findings showed that our model could simultaneously decrease the tension in the $H_0$ and $S_8$ parameters. When we used only CMB data to test our model, we found that the Hubble parameter's value was $70.84\pm 0.42~\rm{km/s/Mpc}$. This outcome reduced the tension to approximately $2.2\sigma$ compared to the value measured by the SH0ES team, which was $73.04\pm1.04~\rm{km/s/Mpc}$~\cite{Riess:2021jrx}. 

We utilized Dark Energy Survey (DES-Y3) data to investigate whether our model could help alleviate the $S_8$ tension. Our findings indicated that the $\phi^{2/3}$ model was able to reduce the tension to about $\sim 0.82 \sigma$ when comparing the $S_8$ parameter values obtained from CMB data analysis with those derived from analyzing only DES data.

\section*{Acknowledgments}
We are grateful to Hassan Firouzjahi, Alireza Talebian and Fereshteh Felegary for useful discussions and correspondence.

\bibliography{references}

\begin{thebibliography}{97}%
\makeatletter
\providecommand \@ifxundefined [1]{%
 \@ifx{#1\undefined}
}%
\providecommand \@ifnum [1]{%
 \ifnum #1\expandafter \@firstoftwo
 \else \expandafter \@secondoftwo
 \fi
}%
\providecommand \@ifx [1]{%
 \ifx #1\expandafter \@firstoftwo
 \else \expandafter \@secondoftwo
 \fi
}%
\providecommand \natexlab [1]{#1}%
\providecommand \enquote  [1]{``#1''}%
\providecommand \bibnamefont  [1]{#1}%
\providecommand \bibfnamefont [1]{#1}%
\providecommand \citenamefont [1]{#1}%
\providecommand \href@noop [0]{\@secondoftwo}%
\providecommand \href [0]{\begingroup \@sanitize@url \@href}%
\providecommand \@href[1]{\@@startlink{#1}\@@href}%
\providecommand \@@href[1]{\endgroup#1\@@endlink}%
\providecommand \@sanitize@url [0]{\catcode `\\12\catcode `\$12\catcode
  `\&12\catcode `\#12\catcode `\^12\catcode `\_12\catcode `\%12\relax}%
\providecommand \@@startlink[1]{}%
\providecommand \@@endlink[0]{}%
\providecommand \url  [0]{\begingroup\@sanitize@url \@url }%
\providecommand \@url [1]{\endgroup\@href {#1}{\urlprefix }}%
\providecommand \urlprefix  [0]{URL }%
\providecommand \Eprint [0]{\href }%
\providecommand \doibase [0]{http://dx.doi.org/}%
\providecommand \selectlanguage [0]{\@gobble}%
\providecommand \bibinfo  [0]{\@secondoftwo}%
\providecommand \bibfield  [0]{\@secondoftwo}%
\providecommand \translation [1]{[#1]}%
\providecommand \BibitemOpen [0]{}%
\providecommand \bibitemStop [0]{}%
\providecommand \bibitemNoStop [0]{.\EOS\space}%
\providecommand \EOS [0]{\spacefactor3000\relax}%
\providecommand \BibitemShut  [1]{\csname bibitem#1\endcsname}%
\let\auto@bib@innerbib\@empty
%</preamble>
\bibitem [{\citenamefont {Riess}\ \emph {et~al.}(1998)\citenamefont {Riess}
  \emph {et~al.}}]{SupernovaSearchTeam:1998fmf}%
  \BibitemOpen
  \bibfield  {author} {\bibinfo {author} {\bibfnamefont {A.~G.}\ \bibnamefont
  {Riess}} \emph {et~al.} (\bibinfo {collaboration} {Supernova Search Team}),\
  }\href {\doibase 10.1086/300499} {\bibfield  {journal} {\bibinfo  {journal}
  {Astron. J.}\ }\textbf {\bibinfo {volume} {116}},\ \bibinfo {pages} {1009}
  (\bibinfo {year} {1998})},\ \Eprint {http://arxiv.org/abs/astro-ph/9805201}
  {arXiv:astro-ph/9805201} \BibitemShut {NoStop}%
\bibitem [{\citenamefont {Perlmutter}\ \emph {et~al.}(1999)\citenamefont
  {Perlmutter} \emph {et~al.}}]{SupernovaCosmologyProject:1998vns}%
  \BibitemOpen
  \bibfield  {author} {\bibinfo {author} {\bibfnamefont {S.}~\bibnamefont
  {Perlmutter}} \emph {et~al.} (\bibinfo {collaboration} {Supernova Cosmology
  Project}),\ }\href {\doibase 10.1086/307221} {\bibfield  {journal} {\bibinfo
  {journal} {Astrophys. J.}\ }\textbf {\bibinfo {volume} {517}},\ \bibinfo
  {pages} {565} (\bibinfo {year} {1999})},\ \Eprint
  {http://arxiv.org/abs/astro-ph/9812133} {arXiv:astro-ph/9812133} \BibitemShut
  {NoStop}%
\bibitem [{\citenamefont {Troxel}\ \emph {et~al.}(2018)\citenamefont {Troxel}
  \emph {et~al.}}]{DES:2017qwj}%
  \BibitemOpen
  \bibfield  {author} {\bibinfo {author} {\bibfnamefont {M.~A.}\ \bibnamefont
  {Troxel}} \emph {et~al.} (\bibinfo {collaboration} {DES}),\ }\href {\doibase
  10.1103/PhysRevD.98.043528} {\bibfield  {journal} {\bibinfo  {journal} {Phys.
  Rev. D}\ }\textbf {\bibinfo {volume} {98}},\ \bibinfo {pages} {043528}
  (\bibinfo {year} {2018})},\ \Eprint {http://arxiv.org/abs/1708.01538}
  {arXiv:1708.01538 [astro-ph.CO]} \BibitemShut {NoStop}%
\bibitem [{\citenamefont {Aghanim}\ \emph
  {et~al.}(2020{\natexlab{a}})\citenamefont {Aghanim} \emph
  {et~al.}}]{Planck:2018vyg}%
  \BibitemOpen
  \bibfield  {author} {\bibinfo {author} {\bibfnamefont {N.}~\bibnamefont
  {Aghanim}} \emph {et~al.} (\bibinfo {collaboration} {Planck}),\ }\href
  {\doibase 10.1051/0004-6361/201833910} {\bibfield  {journal} {\bibinfo
  {journal} {Astron. Astrophys.}\ }\textbf {\bibinfo {volume} {641}},\ \bibinfo
  {pages} {A6} (\bibinfo {year} {2020}{\natexlab{a}})},\ \bibinfo {note}
  {[Erratum: Astron.Astrophys. 652, C4 (2021)]},\ \Eprint
  {http://arxiv.org/abs/1807.06209} {arXiv:1807.06209 [astro-ph.CO]}
  \BibitemShut {NoStop}%
\bibitem [{\citenamefont {Bianchini}\ \emph {et~al.}(2020)\citenamefont
  {Bianchini} \emph {et~al.}}]{SPT:2019fqo}%
  \BibitemOpen
  \bibfield  {author} {\bibinfo {author} {\bibfnamefont {F.}~\bibnamefont
  {Bianchini}} \emph {et~al.} (\bibinfo {collaboration} {SPT}),\ }\href
  {\doibase 10.3847/1538-4357/ab6082} {\bibfield  {journal} {\bibinfo
  {journal} {Astrophys. J.}\ }\textbf {\bibinfo {volume} {888}},\ \bibinfo
  {pages} {119} (\bibinfo {year} {2020})},\ \Eprint
  {http://arxiv.org/abs/1910.07157} {arXiv:1910.07157 [astro-ph.CO]}
  \BibitemShut {NoStop}%
\bibitem [{\citenamefont {Aiola}\ \emph {et~al.}(2020)\citenamefont {Aiola}
  \emph {et~al.}}]{ACT:2020gnv}%
  \BibitemOpen
  \bibfield  {author} {\bibinfo {author} {\bibfnamefont {S.}~\bibnamefont
  {Aiola}} \emph {et~al.} (\bibinfo {collaboration} {ACT}),\ }\href {\doibase
  10.1088/1475-7516/2020/12/047} {\bibfield  {journal} {\bibinfo  {journal}
  {JCAP}\ }\textbf {\bibinfo {volume} {12}},\ \bibinfo {pages} {047} (\bibinfo
  {year} {2020})},\ \Eprint {http://arxiv.org/abs/2007.07288} {arXiv:2007.07288
  [astro-ph.CO]} \BibitemShut {NoStop}%
\bibitem [{\citenamefont {Alam}\ \emph {et~al.}(2021)\citenamefont {Alam} \emph
  {et~al.}}]{eBOSS:2020yzd}%
  \BibitemOpen
  \bibfield  {author} {\bibinfo {author} {\bibfnamefont {S.}~\bibnamefont
  {Alam}} \emph {et~al.} (\bibinfo {collaboration} {eBOSS}),\ }\href {\doibase
  10.1103/PhysRevD.103.083533} {\bibfield  {journal} {\bibinfo  {journal}
  {Phys. Rev. D}\ }\textbf {\bibinfo {volume} {103}},\ \bibinfo {pages}
  {083533} (\bibinfo {year} {2021})},\ \Eprint
  {http://arxiv.org/abs/2007.08991} {arXiv:2007.08991 [astro-ph.CO]}
  \BibitemShut {NoStop}%
\bibitem [{\citenamefont {Asgari}\ \emph {et~al.}(2021)\citenamefont {Asgari}
  \emph {et~al.}}]{KiDS:2020suj}%
  \BibitemOpen
  \bibfield  {author} {\bibinfo {author} {\bibfnamefont {M.}~\bibnamefont
  {Asgari}} \emph {et~al.} (\bibinfo {collaboration} {KiDS}),\ }\href {\doibase
  10.1051/0004-6361/202039070} {\bibfield  {journal} {\bibinfo  {journal}
  {Astron. Astrophys.}\ }\textbf {\bibinfo {volume} {645}},\ \bibinfo {pages}
  {A104} (\bibinfo {year} {2021})},\ \Eprint {http://arxiv.org/abs/2007.15633}
  {arXiv:2007.15633 [astro-ph.CO]} \BibitemShut {NoStop}%
\bibitem [{\citenamefont {Mossa}\ \emph {et~al.}(2020)\citenamefont {Mossa}
  \emph {et~al.}}]{Mossa:2020gjc}%
  \BibitemOpen
  \bibfield  {author} {\bibinfo {author} {\bibfnamefont {V.}~\bibnamefont
  {Mossa}} \emph {et~al.},\ }\href {\doibase 10.1038/s41586-020-2878-4}
  {\bibfield  {journal} {\bibinfo  {journal} {Nature}\ }\textbf {\bibinfo
  {volume} {587}},\ \bibinfo {pages} {210} (\bibinfo {year}
  {2020})}\BibitemShut {NoStop}%
\bibitem [{\citenamefont {Brout}\ \emph {et~al.}(2022)\citenamefont {Brout}
  \emph {et~al.}}]{Brout:2022vxf}%
  \BibitemOpen
  \bibfield  {author} {\bibinfo {author} {\bibfnamefont {D.}~\bibnamefont
  {Brout}} \emph {et~al.},\ }\href {\doibase 10.3847/1538-4357/ac8e04}
  {\bibfield  {journal} {\bibinfo  {journal} {Astrophys. J.}\ }\textbf
  {\bibinfo {volume} {938}},\ \bibinfo {pages} {110} (\bibinfo {year}
  {2022})},\ \Eprint {http://arxiv.org/abs/2202.04077} {arXiv:2202.04077
  [astro-ph.CO]} \BibitemShut {NoStop}%
\bibitem [{\citenamefont {Perivolaropoulos}\ and\ \citenamefont
  {Skara}(2022)}]{Perivolaropoulos:2021jda}%
  \BibitemOpen
  \bibfield  {author} {\bibinfo {author} {\bibfnamefont {L.}~\bibnamefont
  {Perivolaropoulos}}\ and\ \bibinfo {author} {\bibfnamefont {F.}~\bibnamefont
  {Skara}},\ }\href {\doibase 10.1016/j.newar.2022.101659} {\bibfield
  {journal} {\bibinfo  {journal} {New Astron. Rev.}\ }\textbf {\bibinfo
  {volume} {95}},\ \bibinfo {pages} {101659} (\bibinfo {year} {2022})},\
  \Eprint {http://arxiv.org/abs/2105.05208} {arXiv:2105.05208 [astro-ph.CO]}
  \BibitemShut {NoStop}%
\bibitem [{\citenamefont {Huterer}(2023)}]{Huterer:2023qez}%
  \BibitemOpen
  \bibfield  {author} {\bibinfo {author} {\bibfnamefont {D.}~\bibnamefont
  {Huterer}},\ }\href {\doibase 10.1140/epjp/s13360-023-04591-0} {\bibfield
  {journal} {\bibinfo  {journal} {Eur. Phys. J. Plus}\ }\textbf {\bibinfo
  {volume} {138}},\ \bibinfo {pages} {1004} (\bibinfo {year}
  {2023})}\BibitemShut {NoStop}%
\bibitem [{\citenamefont {Riess}\ \emph {et~al.}(2022)\citenamefont {Riess}
  \emph {et~al.}}]{Riess:2021jrx}%
  \BibitemOpen
  \bibfield  {author} {\bibinfo {author} {\bibfnamefont {A.~G.}\ \bibnamefont
  {Riess}} \emph {et~al.},\ }\href {\doibase 10.3847/2041-8213/ac5c5b}
  {\bibfield  {journal} {\bibinfo  {journal} {Astrophys. J. Lett.}\ }\textbf
  {\bibinfo {volume} {934}},\ \bibinfo {pages} {L7} (\bibinfo {year} {2022})},\
  \Eprint {http://arxiv.org/abs/2112.04510} {arXiv:2112.04510 [astro-ph.CO]}
  \BibitemShut {NoStop}%
\bibitem [{\citenamefont {Verde}\ \emph {et~al.}(2019)\citenamefont {Verde},
  \citenamefont {Treu},\ and\ \citenamefont {Riess}}]{Verde:2019ivm}%
  \BibitemOpen
  \bibfield  {author} {\bibinfo {author} {\bibfnamefont {L.}~\bibnamefont
  {Verde}}, \bibinfo {author} {\bibfnamefont {T.}~\bibnamefont {Treu}}, \ and\
  \bibinfo {author} {\bibfnamefont {A.~G.}\ \bibnamefont {Riess}},\ }\href
  {\doibase 10.1038/s41550-019-0902-0} {\bibfield  {journal} {\bibinfo
  {journal} {Nature Astron.}\ }\textbf {\bibinfo {volume} {3}},\ \bibinfo
  {pages} {891} (\bibinfo {year} {2019})},\ \Eprint
  {http://arxiv.org/abs/1907.10625} {arXiv:1907.10625 [astro-ph.CO]}
  \BibitemShut {NoStop}%
\bibitem [{\citenamefont {Heymans}\ \emph {et~al.}(2021)\citenamefont {Heymans}
  \emph {et~al.}}]{Heymans:2020gsg}%
  \BibitemOpen
  \bibfield  {author} {\bibinfo {author} {\bibfnamefont {C.}~\bibnamefont
  {Heymans}} \emph {et~al.},\ }\href {\doibase 10.1051/0004-6361/202039063}
  {\bibfield  {journal} {\bibinfo  {journal} {Astron. Astrophys.}\ }\textbf
  {\bibinfo {volume} {646}},\ \bibinfo {pages} {A140} (\bibinfo {year}
  {2021})},\ \Eprint {http://arxiv.org/abs/2007.15632} {arXiv:2007.15632
  [astro-ph.CO]} \BibitemShut {NoStop}%
\bibitem [{\citenamefont {Abbott}\ \emph {et~al.}(2022)\citenamefont {Abbott}
  \emph {et~al.}}]{DES:2021wwk}%
  \BibitemOpen
  \bibfield  {author} {\bibinfo {author} {\bibfnamefont {T.~M.~C.}\
  \bibnamefont {Abbott}} \emph {et~al.} (\bibinfo {collaboration} {DES}),\
  }\href {\doibase 10.1103/PhysRevD.105.023520} {\bibfield  {journal} {\bibinfo
   {journal} {Phys. Rev. D}\ }\textbf {\bibinfo {volume} {105}},\ \bibinfo
  {pages} {023520} (\bibinfo {year} {2022})},\ \Eprint
  {http://arxiv.org/abs/2105.13549} {arXiv:2105.13549 [astro-ph.CO]}
  \BibitemShut {NoStop}%
\bibitem [{\citenamefont {Riess}(2019)}]{Riess:2019qba}%
  \BibitemOpen
  \bibfield  {author} {\bibinfo {author} {\bibfnamefont {A.~G.}\ \bibnamefont
  {Riess}},\ }\href {\doibase 10.1038/s42254-019-0137-0} {\bibfield  {journal}
  {\bibinfo  {journal} {Nature Rev. Phys.}\ }\textbf {\bibinfo {volume} {2}},\
  \bibinfo {pages} {10} (\bibinfo {year} {2019})},\ \Eprint
  {http://arxiv.org/abs/2001.03624} {arXiv:2001.03624 [astro-ph.CO]}
  \BibitemShut {NoStop}%
\bibitem [{\citenamefont {Di~Valentino}\ \emph
  {et~al.}(2021{\natexlab{a}})\citenamefont {Di~Valentino} \emph
  {et~al.}}]{DiValentino:2020zio}%
  \BibitemOpen
  \bibfield  {author} {\bibinfo {author} {\bibfnamefont {E.}~\bibnamefont
  {Di~Valentino}} \emph {et~al.},\ }\href {\doibase
  10.1016/j.astropartphys.2021.102605} {\bibfield  {journal} {\bibinfo
  {journal} {Astropart. Phys.}\ }\textbf {\bibinfo {volume} {131}},\ \bibinfo
  {pages} {102605} (\bibinfo {year} {2021}{\natexlab{a}})},\ \Eprint
  {http://arxiv.org/abs/2008.11284} {arXiv:2008.11284 [astro-ph.CO]}
  \BibitemShut {NoStop}%
\bibitem [{\citenamefont {Di~Valentino}\ \emph
  {et~al.}(2021{\natexlab{b}})\citenamefont {Di~Valentino}, \citenamefont
  {Mena}, \citenamefont {Pan}, \citenamefont {Visinelli}, \citenamefont {Yang},
  \citenamefont {Melchiorri}, \citenamefont {Mota}, \citenamefont {Riess},\
  and\ \citenamefont {Silk}}]{DiValentino:2021izs}%
  \BibitemOpen
  \bibfield  {author} {\bibinfo {author} {\bibfnamefont {E.}~\bibnamefont
  {Di~Valentino}}, \bibinfo {author} {\bibfnamefont {O.}~\bibnamefont {Mena}},
  \bibinfo {author} {\bibfnamefont {S.}~\bibnamefont {Pan}}, \bibinfo {author}
  {\bibfnamefont {L.}~\bibnamefont {Visinelli}}, \bibinfo {author}
  {\bibfnamefont {W.}~\bibnamefont {Yang}}, \bibinfo {author} {\bibfnamefont
  {A.}~\bibnamefont {Melchiorri}}, \bibinfo {author} {\bibfnamefont {D.~F.}\
  \bibnamefont {Mota}}, \bibinfo {author} {\bibfnamefont {A.~G.}\ \bibnamefont
  {Riess}}, \ and\ \bibinfo {author} {\bibfnamefont {J.}~\bibnamefont {Silk}},\
  }\href {\doibase 10.1088/1361-6382/ac086d} {\bibfield  {journal} {\bibinfo
  {journal} {Class. Quant. Grav.}\ }\textbf {\bibinfo {volume} {38}},\ \bibinfo
  {pages} {153001} (\bibinfo {year} {2021}{\natexlab{b}})},\ \Eprint
  {http://arxiv.org/abs/2103.01183} {arXiv:2103.01183 [astro-ph.CO]}
  \BibitemShut {NoStop}%
\bibitem [{\citenamefont {Sch\"oneberg}\ \emph {et~al.}(2022)\citenamefont
  {Sch\"oneberg}, \citenamefont {Franco~Abell\'an}, \citenamefont
  {P\'erez~S\'anchez}, \citenamefont {Witte}, \citenamefont {Poulin},\ and\
  \citenamefont {Lesgourgues}}]{Schoneberg:2021qvd}%
  \BibitemOpen
  \bibfield  {author} {\bibinfo {author} {\bibfnamefont {N.}~\bibnamefont
  {Sch\"oneberg}}, \bibinfo {author} {\bibfnamefont {G.}~\bibnamefont
  {Franco~Abell\'an}}, \bibinfo {author} {\bibfnamefont {A.}~\bibnamefont
  {P\'erez~S\'anchez}}, \bibinfo {author} {\bibfnamefont {S.~J.}\ \bibnamefont
  {Witte}}, \bibinfo {author} {\bibfnamefont {V.}~\bibnamefont {Poulin}}, \
  and\ \bibinfo {author} {\bibfnamefont {J.}~\bibnamefont {Lesgourgues}},\
  }\href {\doibase 10.1016/j.physrep.2022.07.001} {\bibfield  {journal}
  {\bibinfo  {journal} {Phys. Rept.}\ }\textbf {\bibinfo {volume} {984}},\
  \bibinfo {pages} {1} (\bibinfo {year} {2022})},\ \Eprint
  {http://arxiv.org/abs/2107.10291} {arXiv:2107.10291 [astro-ph.CO]}
  \BibitemShut {NoStop}%
\bibitem [{\citenamefont {Shah}\ \emph {et~al.}(2021)\citenamefont {Shah},
  \citenamefont {Lemos},\ and\ \citenamefont {Lahav}}]{Shah:2021onj}%
  \BibitemOpen
  \bibfield  {author} {\bibinfo {author} {\bibfnamefont {P.}~\bibnamefont
  {Shah}}, \bibinfo {author} {\bibfnamefont {P.}~\bibnamefont {Lemos}}, \ and\
  \bibinfo {author} {\bibfnamefont {O.}~\bibnamefont {Lahav}},\ }\href
  {\doibase 10.1007/s00159-021-00137-4} {\bibfield  {journal} {\bibinfo
  {journal} {Astron. Astrophys. Rev.}\ }\textbf {\bibinfo {volume} {29}},\
  \bibinfo {pages} {9} (\bibinfo {year} {2021})},\ \Eprint
  {http://arxiv.org/abs/2109.01161} {arXiv:2109.01161 [astro-ph.CO]}
  \BibitemShut {NoStop}%
\bibitem [{\citenamefont {Abdalla}\ \emph {et~al.}(2022)\citenamefont {Abdalla}
  \emph {et~al.}}]{Abdalla:2022yfr}%
  \BibitemOpen
  \bibfield  {author} {\bibinfo {author} {\bibfnamefont {E.}~\bibnamefont
  {Abdalla}} \emph {et~al.},\ }\href {\doibase 10.1016/j.jheap.2022.04.002}
  {\bibfield  {journal} {\bibinfo  {journal} {JHEAp}\ }\textbf {\bibinfo
  {volume} {34}},\ \bibinfo {pages} {49} (\bibinfo {year} {2022})},\ \Eprint
  {http://arxiv.org/abs/2203.06142} {arXiv:2203.06142 [astro-ph.CO]}
  \BibitemShut {NoStop}%
\bibitem [{\citenamefont {Di~Valentino}(2022)}]{DiValentino:2022fjm}%
  \BibitemOpen
  \bibfield  {author} {\bibinfo {author} {\bibfnamefont {E.}~\bibnamefont
  {Di~Valentino}},\ }\href {\doibase 10.3390/universe8080399} {\bibfield
  {journal} {\bibinfo  {journal} {Universe}\ }\textbf {\bibinfo {volume} {8}},\
  \bibinfo {pages} {399} (\bibinfo {year} {2022})}\BibitemShut {NoStop}%
\bibitem [{\citenamefont {Hu}\ and\ \citenamefont {Wang}(2023)}]{Hu:2023jqc}%
  \BibitemOpen
  \bibfield  {author} {\bibinfo {author} {\bibfnamefont {J.-P.}\ \bibnamefont
  {Hu}}\ and\ \bibinfo {author} {\bibfnamefont {F.-Y.}\ \bibnamefont {Wang}},\
  }\href {\doibase 10.3390/universe9020094} {\bibfield  {journal} {\bibinfo
  {journal} {Universe}\ }\textbf {\bibinfo {volume} {9}},\ \bibinfo {pages}
  {94} (\bibinfo {year} {2023})},\ \Eprint {http://arxiv.org/abs/2302.05709}
  {arXiv:2302.05709 [astro-ph.CO]} \BibitemShut {NoStop}%
\bibitem [{\citenamefont {Heisenberg}\ \emph {et~al.}(2023)\citenamefont
  {Heisenberg}, \citenamefont {Villarrubia-Rojo},\ and\ \citenamefont
  {Zosso}}]{Heisenberg:2022lob}%
  \BibitemOpen
  \bibfield  {author} {\bibinfo {author} {\bibfnamefont {L.}~\bibnamefont
  {Heisenberg}}, \bibinfo {author} {\bibfnamefont {H.}~\bibnamefont
  {Villarrubia-Rojo}}, \ and\ \bibinfo {author} {\bibfnamefont
  {J.}~\bibnamefont {Zosso}},\ }\href {\doibase 10.1016/j.dark.2022.101163}
  {\bibfield  {journal} {\bibinfo  {journal} {Phys. Dark Univ.}\ }\textbf
  {\bibinfo {volume} {39}},\ \bibinfo {pages} {101163} (\bibinfo {year}
  {2023})},\ \Eprint {http://arxiv.org/abs/2201.11623} {arXiv:2201.11623
  [astro-ph.CO]} \BibitemShut {NoStop}%
\bibitem [{\citenamefont {Heisenberg}\ \emph {et~al.}(2022)\citenamefont
  {Heisenberg}, \citenamefont {Villarrubia-Rojo},\ and\ \citenamefont
  {Zosso}}]{Heisenberg:2022gqk}%
  \BibitemOpen
  \bibfield  {author} {\bibinfo {author} {\bibfnamefont {L.}~\bibnamefont
  {Heisenberg}}, \bibinfo {author} {\bibfnamefont {H.}~\bibnamefont
  {Villarrubia-Rojo}}, \ and\ \bibinfo {author} {\bibfnamefont
  {J.}~\bibnamefont {Zosso}},\ }\href {\doibase 10.1103/PhysRevD.106.043503}
  {\bibfield  {journal} {\bibinfo  {journal} {Phys. Rev. D}\ }\textbf {\bibinfo
  {volume} {106}},\ \bibinfo {pages} {043503} (\bibinfo {year} {2022})},\
  \Eprint {http://arxiv.org/abs/2202.01202} {arXiv:2202.01202 [astro-ph.CO]}
  \BibitemShut {NoStop}%
\bibitem [{\citenamefont {Braglia}\ \emph {et~al.}(2020)\citenamefont
  {Braglia}, \citenamefont {Ballardini}, \citenamefont {Emond}, \citenamefont
  {Finelli}, \citenamefont {Gumrukcuoglu}, \citenamefont {Koyama},\ and\
  \citenamefont {Paoletti}}]{Braglia:2020iik}%
  \BibitemOpen
  \bibfield  {author} {\bibinfo {author} {\bibfnamefont {M.}~\bibnamefont
  {Braglia}}, \bibinfo {author} {\bibfnamefont {M.}~\bibnamefont {Ballardini}},
  \bibinfo {author} {\bibfnamefont {W.~T.}\ \bibnamefont {Emond}}, \bibinfo
  {author} {\bibfnamefont {F.}~\bibnamefont {Finelli}}, \bibinfo {author}
  {\bibfnamefont {A.~E.}\ \bibnamefont {Gumrukcuoglu}}, \bibinfo {author}
  {\bibfnamefont {K.}~\bibnamefont {Koyama}}, \ and\ \bibinfo {author}
  {\bibfnamefont {D.}~\bibnamefont {Paoletti}},\ }\href {\doibase
  10.1103/PhysRevD.102.023529} {\bibfield  {journal} {\bibinfo  {journal}
  {Phys. Rev. D}\ }\textbf {\bibinfo {volume} {102}},\ \bibinfo {pages}
  {023529} (\bibinfo {year} {2020})},\ \Eprint
  {http://arxiv.org/abs/2004.11161} {arXiv:2004.11161 [astro-ph.CO]}
  \BibitemShut {NoStop}%
\bibitem [{\citenamefont {Lee}\ \emph {et~al.}(2022)\citenamefont {Lee},
  \citenamefont {Lee}, \citenamefont {Colg\'ain}, \citenamefont
  {Sheikh-Jabbari},\ and\ \citenamefont {Thakur}}]{Lee:2022cyh}%
  \BibitemOpen
  \bibfield  {author} {\bibinfo {author} {\bibfnamefont {B.-H.}\ \bibnamefont
  {Lee}}, \bibinfo {author} {\bibfnamefont {W.}~\bibnamefont {Lee}}, \bibinfo
  {author} {\bibfnamefont {E.~O.}\ \bibnamefont {Colg\'ain}}, \bibinfo {author}
  {\bibfnamefont {M.~M.}\ \bibnamefont {Sheikh-Jabbari}}, \ and\ \bibinfo
  {author} {\bibfnamefont {S.}~\bibnamefont {Thakur}},\ }\href {\doibase
  10.1088/1475-7516/2022/04/004} {\bibfield  {journal} {\bibinfo  {journal}
  {JCAP}\ }\textbf {\bibinfo {volume} {04}},\ \bibinfo {pages} {004} (\bibinfo
  {year} {2022})},\ \Eprint {http://arxiv.org/abs/2202.03906} {arXiv:2202.03906
  [astro-ph.CO]} \BibitemShut {NoStop}%
\bibitem [{\citenamefont {Krishnan}\ \emph {et~al.}(2020)\citenamefont
  {Krishnan}, \citenamefont {Colg\'ain}, \citenamefont {Ruchika}, \citenamefont
  {Sen}, \citenamefont {Sheikh-Jabbari},\ and\ \citenamefont
  {Yang}}]{Krishnan:2020obg}%
  \BibitemOpen
  \bibfield  {author} {\bibinfo {author} {\bibfnamefont {C.}~\bibnamefont
  {Krishnan}}, \bibinfo {author} {\bibfnamefont {E.~O.}\ \bibnamefont
  {Colg\'ain}}, \bibinfo {author} {\bibnamefont {Ruchika}}, \bibinfo {author}
  {\bibfnamefont {A.~A.}\ \bibnamefont {Sen}}, \bibinfo {author} {\bibfnamefont
  {M.~M.}\ \bibnamefont {Sheikh-Jabbari}}, \ and\ \bibinfo {author}
  {\bibfnamefont {T.}~\bibnamefont {Yang}},\ }\href {\doibase
  10.1103/PhysRevD.102.103525} {\bibfield  {journal} {\bibinfo  {journal}
  {Phys. Rev. D}\ }\textbf {\bibinfo {volume} {102}},\ \bibinfo {pages}
  {103525} (\bibinfo {year} {2020})},\ \Eprint
  {http://arxiv.org/abs/2002.06044} {arXiv:2002.06044 [astro-ph.CO]}
  \BibitemShut {NoStop}%
\bibitem [{\citenamefont {Bernal}\ \emph {et~al.}(2016)\citenamefont {Bernal},
  \citenamefont {Verde},\ and\ \citenamefont {Riess}}]{Bernal:2016gxb}%
  \BibitemOpen
  \bibfield  {author} {\bibinfo {author} {\bibfnamefont {J.~L.}\ \bibnamefont
  {Bernal}}, \bibinfo {author} {\bibfnamefont {L.}~\bibnamefont {Verde}}, \
  and\ \bibinfo {author} {\bibfnamefont {A.~G.}\ \bibnamefont {Riess}},\ }\href
  {\doibase 10.1088/1475-7516/2016/10/019} {\bibfield  {journal} {\bibinfo
  {journal} {JCAP}\ }\textbf {\bibinfo {volume} {10}},\ \bibinfo {pages} {019}
  (\bibinfo {year} {2016})},\ \Eprint {http://arxiv.org/abs/1607.05617}
  {arXiv:1607.05617 [astro-ph.CO]} \BibitemShut {NoStop}%
\bibitem [{\citenamefont {Vagnozzi}(2020)}]{Vagnozzi:2019ezj}%
  \BibitemOpen
  \bibfield  {author} {\bibinfo {author} {\bibfnamefont {S.}~\bibnamefont
  {Vagnozzi}},\ }\href {\doibase 10.1103/PhysRevD.102.023518} {\bibfield
  {journal} {\bibinfo  {journal} {Phys. Rev. D}\ }\textbf {\bibinfo {volume}
  {102}},\ \bibinfo {pages} {023518} (\bibinfo {year} {2020})},\ \Eprint
  {http://arxiv.org/abs/1907.07569} {arXiv:1907.07569 [astro-ph.CO]}
  \BibitemShut {NoStop}%
\bibitem [{\citenamefont {Guo}\ \emph {et~al.}(2019)\citenamefont {Guo},
  \citenamefont {Zhang},\ and\ \citenamefont {Zhang}}]{Guo:2018ans}%
  \BibitemOpen
  \bibfield  {author} {\bibinfo {author} {\bibfnamefont {R.-Y.}\ \bibnamefont
  {Guo}}, \bibinfo {author} {\bibfnamefont {J.-F.}\ \bibnamefont {Zhang}}, \
  and\ \bibinfo {author} {\bibfnamefont {X.}~\bibnamefont {Zhang}},\ }\href
  {\doibase 10.1088/1475-7516/2019/02/054} {\bibfield  {journal} {\bibinfo
  {journal} {JCAP}\ }\textbf {\bibinfo {volume} {02}},\ \bibinfo {pages} {054}
  (\bibinfo {year} {2019})},\ \Eprint {http://arxiv.org/abs/1809.02340}
  {arXiv:1809.02340 [astro-ph.CO]} \BibitemShut {NoStop}%
\bibitem [{\citenamefont {M\"ortsell}\ and\ \citenamefont
  {Dhawan}(2018)}]{Mortsell:2018mfj}%
  \BibitemOpen
  \bibfield  {author} {\bibinfo {author} {\bibfnamefont {E.}~\bibnamefont
  {M\"ortsell}}\ and\ \bibinfo {author} {\bibfnamefont {S.}~\bibnamefont
  {Dhawan}},\ }\href {\doibase 10.1088/1475-7516/2018/09/025} {\bibfield
  {journal} {\bibinfo  {journal} {JCAP}\ }\textbf {\bibinfo {volume} {09}},\
  \bibinfo {pages} {025} (\bibinfo {year} {2018})},\ \Eprint
  {http://arxiv.org/abs/1801.07260} {arXiv:1801.07260 [astro-ph.CO]}
  \BibitemShut {NoStop}%
\bibitem [{\citenamefont {Mortsell}\ \emph {et~al.}(2021)\citenamefont
  {Mortsell}, \citenamefont {Goobar}, \citenamefont {Johansson},\ and\
  \citenamefont {Dhawan}}]{Mortsell:2021nzg}%
  \BibitemOpen
  \bibfield  {author} {\bibinfo {author} {\bibfnamefont {E.}~\bibnamefont
  {Mortsell}}, \bibinfo {author} {\bibfnamefont {A.}~\bibnamefont {Goobar}},
  \bibinfo {author} {\bibfnamefont {J.}~\bibnamefont {Johansson}}, \ and\
  \bibinfo {author} {\bibfnamefont {S.}~\bibnamefont {Dhawan}},\ }\href@noop {}
  {\  (\bibinfo {year} {2021})},\ \Eprint {http://arxiv.org/abs/2105.11461}
  {arXiv:2105.11461 [astro-ph.CO]} \BibitemShut {NoStop}%
\bibitem [{\citenamefont {Efstathiou}(2020)}]{Efstathiou:2020wxn}%
  \BibitemOpen
  \bibfield  {author} {\bibinfo {author} {\bibfnamefont {G.}~\bibnamefont
  {Efstathiou}},\ }\href@noop {} {\  (\bibinfo {year} {2020})},\ \Eprint
  {http://arxiv.org/abs/2007.10716} {arXiv:2007.10716 [astro-ph.CO]}
  \BibitemShut {NoStop}%
\bibitem [{\citenamefont {Petronikolou}\ \emph {et~al.}(2022)\citenamefont
  {Petronikolou}, \citenamefont {Basilakos},\ and\ \citenamefont
  {Saridakis}}]{Petronikolou:2021shp}%
  \BibitemOpen
  \bibfield  {author} {\bibinfo {author} {\bibfnamefont {M.}~\bibnamefont
  {Petronikolou}}, \bibinfo {author} {\bibfnamefont {S.}~\bibnamefont
  {Basilakos}}, \ and\ \bibinfo {author} {\bibfnamefont {E.~N.}\ \bibnamefont
  {Saridakis}},\ }\href {\doibase 10.1103/PhysRevD.106.124051} {\bibfield
  {journal} {\bibinfo  {journal} {Phys. Rev. D}\ }\textbf {\bibinfo {volume}
  {106}},\ \bibinfo {pages} {124051} (\bibinfo {year} {2022})},\ \Eprint
  {http://arxiv.org/abs/2110.01338} {arXiv:2110.01338 [gr-qc]} \BibitemShut
  {NoStop}%
\bibitem [{\citenamefont {Basilakos}\ \emph {et~al.}(2024)\citenamefont
  {Basilakos}, \citenamefont {Lymperis}, \citenamefont {Petronikolou},\ and\
  \citenamefont {Saridakis}}]{Basilakos:2023kvk}%
  \BibitemOpen
  \bibfield  {author} {\bibinfo {author} {\bibfnamefont {S.}~\bibnamefont
  {Basilakos}}, \bibinfo {author} {\bibfnamefont {A.}~\bibnamefont {Lymperis}},
  \bibinfo {author} {\bibfnamefont {M.}~\bibnamefont {Petronikolou}}, \ and\
  \bibinfo {author} {\bibfnamefont {E.~N.}\ \bibnamefont {Saridakis}},\ }\href
  {\doibase 10.1140/epjc/s10052-024-12573-4} {\bibfield  {journal} {\bibinfo
  {journal} {Eur. Phys. J. C}\ }\textbf {\bibinfo {volume} {84}},\ \bibinfo
  {pages} {297} (\bibinfo {year} {2024})},\ \Eprint
  {http://arxiv.org/abs/2308.01200} {arXiv:2308.01200 [gr-qc]} \BibitemShut
  {NoStop}%
\bibitem [{\citenamefont {Gangopadhyay}\ \emph {et~al.}(2023)\citenamefont
  {Gangopadhyay}, \citenamefont {Sami},\ and\ \citenamefont
  {Sharma}}]{Gangopadhyay:2023nli}%
  \BibitemOpen
  \bibfield  {author} {\bibinfo {author} {\bibfnamefont {M.~R.}\ \bibnamefont
  {Gangopadhyay}}, \bibinfo {author} {\bibfnamefont {M.}~\bibnamefont {Sami}},
  \ and\ \bibinfo {author} {\bibfnamefont {M.~K.}\ \bibnamefont {Sharma}},\
  }\href {\doibase 10.1103/PhysRevD.108.103526} {\bibfield  {journal} {\bibinfo
   {journal} {Phys. Rev. D}\ }\textbf {\bibinfo {volume} {108}},\ \bibinfo
  {pages} {103526} (\bibinfo {year} {2023})},\ \Eprint
  {http://arxiv.org/abs/2303.07301} {arXiv:2303.07301 [astro-ph.CO]}
  \BibitemShut {NoStop}%
\bibitem [{\citenamefont {Sharma}\ \emph {et~al.}(2023)\citenamefont {Sharma},
  \citenamefont {Pacif}, \citenamefont {Yergaliyeva},\ and\ \citenamefont
  {Yesmakhanova}}]{Sharma:2022oxh}%
  \BibitemOpen
  \bibfield  {author} {\bibinfo {author} {\bibfnamefont {M.~K.}\ \bibnamefont
  {Sharma}}, \bibinfo {author} {\bibfnamefont {S.~K.~J.}\ \bibnamefont
  {Pacif}}, \bibinfo {author} {\bibfnamefont {G.}~\bibnamefont {Yergaliyeva}},
  \ and\ \bibinfo {author} {\bibfnamefont {K.}~\bibnamefont {Yesmakhanova}},\
  }\href {\doibase 10.1016/j.aop.2023.169345} {\bibfield  {journal} {\bibinfo
  {journal} {Annals Phys.}\ }\textbf {\bibinfo {volume} {454}},\ \bibinfo
  {pages} {169345} (\bibinfo {year} {2023})},\ \Eprint
  {http://arxiv.org/abs/2205.13514} {arXiv:2205.13514 [gr-qc]} \BibitemShut
  {NoStop}%
\bibitem [{\citenamefont {Moshafi}\ \emph {et~al.}(2022)\citenamefont
  {Moshafi}, \citenamefont {Firouzjahi},\ and\ \citenamefont
  {Talebian}}]{Moshafi:2022mva}%
  \BibitemOpen
  \bibfield  {author} {\bibinfo {author} {\bibfnamefont {H.}~\bibnamefont
  {Moshafi}}, \bibinfo {author} {\bibfnamefont {H.}~\bibnamefont {Firouzjahi}},
  \ and\ \bibinfo {author} {\bibfnamefont {A.}~\bibnamefont {Talebian}},\
  }\href {\doibase 10.3847/1538-4357/ac9c58} {\bibfield  {journal} {\bibinfo
  {journal} {Astrophys. J.}\ }\textbf {\bibinfo {volume} {940}},\ \bibinfo
  {pages} {121} (\bibinfo {year} {2022})},\ \Eprint
  {http://arxiv.org/abs/2208.05583} {arXiv:2208.05583 [astro-ph.CO]}
  \BibitemShut {NoStop}%
\bibitem [{\citenamefont {Poulin}\ \emph {et~al.}(2019)\citenamefont {Poulin},
  \citenamefont {Smith}, \citenamefont {Karwal},\ and\ \citenamefont
  {Kamionkowski}}]{Poulin:2018cxd}%
  \BibitemOpen
  \bibfield  {author} {\bibinfo {author} {\bibfnamefont {V.}~\bibnamefont
  {Poulin}}, \bibinfo {author} {\bibfnamefont {T.~L.}\ \bibnamefont {Smith}},
  \bibinfo {author} {\bibfnamefont {T.}~\bibnamefont {Karwal}}, \ and\ \bibinfo
  {author} {\bibfnamefont {M.}~\bibnamefont {Kamionkowski}},\ }\href {\doibase
  10.1103/PhysRevLett.122.221301} {\bibfield  {journal} {\bibinfo  {journal}
  {Phys. Rev. Lett.}\ }\textbf {\bibinfo {volume} {122}},\ \bibinfo {pages}
  {221301} (\bibinfo {year} {2019})},\ \Eprint
  {http://arxiv.org/abs/1811.04083} {arXiv:1811.04083 [astro-ph.CO]}
  \BibitemShut {NoStop}%
\bibitem [{\citenamefont {Elizalde}\ \emph {et~al.}(2020)\citenamefont
  {Elizalde}, \citenamefont {Khurshudyan}, \citenamefont {Odintsov},\ and\
  \citenamefont {Myrzakulov}}]{Elizalde:2020mfs}%
  \BibitemOpen
  \bibfield  {author} {\bibinfo {author} {\bibfnamefont {E.}~\bibnamefont
  {Elizalde}}, \bibinfo {author} {\bibfnamefont {M.}~\bibnamefont
  {Khurshudyan}}, \bibinfo {author} {\bibfnamefont {S.~D.}\ \bibnamefont
  {Odintsov}}, \ and\ \bibinfo {author} {\bibfnamefont {R.}~\bibnamefont
  {Myrzakulov}},\ }\href {\doibase 10.1103/PhysRevD.102.123501} {\bibfield
  {journal} {\bibinfo  {journal} {Phys. Rev. D}\ }\textbf {\bibinfo {volume}
  {102}},\ \bibinfo {pages} {123501} (\bibinfo {year} {2020})},\ \Eprint
  {http://arxiv.org/abs/2006.01879} {arXiv:2006.01879 [gr-qc]} \BibitemShut
  {NoStop}%
\bibitem [{\citenamefont {Mostaghel}\ \emph {et~al.}(2018)\citenamefont
  {Mostaghel}, \citenamefont {Moshafi},\ and\ \citenamefont
  {Movahed}}]{Mostaghel:2018pia}%
  \BibitemOpen
  \bibfield  {author} {\bibinfo {author} {\bibfnamefont {B.}~\bibnamefont
  {Mostaghel}}, \bibinfo {author} {\bibfnamefont {H.}~\bibnamefont {Moshafi}},
  \ and\ \bibinfo {author} {\bibfnamefont {S.~M.~S.}\ \bibnamefont {Movahed}},\
  }\href {\doibase 10.1093/mnras/sty2384} {\bibfield  {journal} {\bibinfo
  {journal} {Mon. Not. Roy. Astron. Soc.}\ }\textbf {\bibinfo {volume} {481}},\
  \bibinfo {pages} {1799} (\bibinfo {year} {2018})},\ \Eprint
  {http://arxiv.org/abs/1810.04856} {arXiv:1810.04856 [astro-ph.CO]}
  \BibitemShut {NoStop}%
\bibitem [{\citenamefont {Mostaghel}\ \emph {et~al.}(2017)\citenamefont
  {Mostaghel}, \citenamefont {Moshafi},\ and\ \citenamefont
  {Movahed}}]{Mostaghel:2016lcd}%
  \BibitemOpen
  \bibfield  {author} {\bibinfo {author} {\bibfnamefont {B.}~\bibnamefont
  {Mostaghel}}, \bibinfo {author} {\bibfnamefont {H.}~\bibnamefont {Moshafi}},
  \ and\ \bibinfo {author} {\bibfnamefont {S.~M.~S.}\ \bibnamefont {Movahed}},\
  }\href {\doibase 10.1140/epjc/s10052-017-5085-1} {\bibfield  {journal}
  {\bibinfo  {journal} {Eur. Phys. J. C}\ }\textbf {\bibinfo {volume} {77}},\
  \bibinfo {pages} {541} (\bibinfo {year} {2017})},\ \Eprint
  {http://arxiv.org/abs/1611.08196} {arXiv:1611.08196 [astro-ph.CO]}
  \BibitemShut {NoStop}%
\bibitem [{\citenamefont {De~Felice}\ \emph {et~al.}(2021)\citenamefont
  {De~Felice}, \citenamefont {Mukohyama},\ and\ \citenamefont
  {Pookkillath}}]{DeFelice:2020cpt}%
  \BibitemOpen
  \bibfield  {author} {\bibinfo {author} {\bibfnamefont {A.}~\bibnamefont
  {De~Felice}}, \bibinfo {author} {\bibfnamefont {S.}~\bibnamefont
  {Mukohyama}}, \ and\ \bibinfo {author} {\bibfnamefont {M.~C.}\ \bibnamefont
  {Pookkillath}},\ }\href {\doibase 10.1016/j.physletb.2021.136201} {\bibfield
  {journal} {\bibinfo  {journal} {Phys. Lett. B}\ }\textbf {\bibinfo {volume}
  {816}},\ \bibinfo {pages} {136201} (\bibinfo {year} {2021})},\ \bibinfo
  {note} {[Erratum: Phys.Lett.B 818, 136364 (2021)]},\ \Eprint
  {http://arxiv.org/abs/2009.08718} {arXiv:2009.08718 [astro-ph.CO]}
  \BibitemShut {NoStop}%
\bibitem [{\citenamefont {Di~Valentino}\ \emph {et~al.}(2024)\citenamefont
  {Di~Valentino}, \citenamefont {Perivolaropoulos},\ and\ \citenamefont
  {Levi~Said}}]{DiValentino:2024wgi}%
  \BibitemOpen
  \bibfield  {author} {\bibinfo {author} {\bibfnamefont {E.}~\bibnamefont
  {Di~Valentino}}, \bibinfo {author} {\bibfnamefont {L.}~\bibnamefont
  {Perivolaropoulos}}, \ and\ \bibinfo {author} {\bibfnamefont
  {J.}~\bibnamefont {Levi~Said}},\ }\href {\doibase 10.3390/universe10040184}
  {\  (\bibinfo {year} {2024}),\ 10.3390/universe10040184},\ \Eprint
  {http://arxiv.org/abs/2404.13981} {arXiv:2404.13981 [gr-qc]} \BibitemShut
  {NoStop}%
\bibitem [{\citenamefont {Di~Valentino}\ \emph
  {et~al.}(2021{\natexlab{c}})\citenamefont {Di~Valentino} \emph
  {et~al.}}]{DiValentino:2020vvd}%
  \BibitemOpen
  \bibfield  {author} {\bibinfo {author} {\bibfnamefont {E.}~\bibnamefont
  {Di~Valentino}} \emph {et~al.},\ }\href {\doibase
  10.1016/j.astropartphys.2021.102604} {\bibfield  {journal} {\bibinfo
  {journal} {Astropart. Phys.}\ }\textbf {\bibinfo {volume} {131}},\ \bibinfo
  {pages} {102604} (\bibinfo {year} {2021}{\natexlab{c}})},\ \Eprint
  {http://arxiv.org/abs/2008.11285} {arXiv:2008.11285 [astro-ph.CO]}
  \BibitemShut {NoStop}%
\bibitem [{\citenamefont {Benisty}(2021)}]{Benisty:2020kdt}%
  \BibitemOpen
  \bibfield  {author} {\bibinfo {author} {\bibfnamefont {D.}~\bibnamefont
  {Benisty}},\ }\href {\doibase 10.1016/j.dark.2020.100766} {\bibfield
  {journal} {\bibinfo  {journal} {Phys. Dark Univ.}\ }\textbf {\bibinfo
  {volume} {31}},\ \bibinfo {pages} {100766} (\bibinfo {year} {2021})},\
  \Eprint {http://arxiv.org/abs/2005.03751} {arXiv:2005.03751 [astro-ph.CO]}
  \BibitemShut {NoStop}%
\bibitem [{\citenamefont {Adi}\ and\ \citenamefont
  {Kovetz}(2021)}]{Adi:2020qqf}%
  \BibitemOpen
  \bibfield  {author} {\bibinfo {author} {\bibfnamefont {T.}~\bibnamefont
  {Adi}}\ and\ \bibinfo {author} {\bibfnamefont {E.~D.}\ \bibnamefont
  {Kovetz}},\ }\href {\doibase 10.1103/PhysRevD.103.023530} {\bibfield
  {journal} {\bibinfo  {journal} {Phys. Rev. D}\ }\textbf {\bibinfo {volume}
  {103}},\ \bibinfo {pages} {023530} (\bibinfo {year} {2021})},\ \Eprint
  {http://arxiv.org/abs/2011.13853} {arXiv:2011.13853 [astro-ph.CO]}
  \BibitemShut {NoStop}%
\bibitem [{\citenamefont {Kazantzidis}\ and\ \citenamefont
  {Perivolaropoulos}(2019)}]{Kazantzidis:2019nuh}%
  \BibitemOpen
  \bibfield  {author} {\bibinfo {author} {\bibfnamefont {L.}~\bibnamefont
  {Kazantzidis}}\ and\ \bibinfo {author} {\bibfnamefont {L.}~\bibnamefont
  {Perivolaropoulos}},\ }\href {\doibase 10.48550/arXiv.1907.03176} {\
  (\bibinfo {year} {2019}),\ 10.48550/arXiv.1907.03176},\ \Eprint
  {http://arxiv.org/abs/1907.03176} {arXiv:1907.03176 [astro-ph.CO]}
  \BibitemShut {NoStop}%
\bibitem [{\citenamefont {Nunes}(2018)}]{Nunes:2018xbm}%
  \BibitemOpen
  \bibfield  {author} {\bibinfo {author} {\bibfnamefont {R.~C.}\ \bibnamefont
  {Nunes}},\ }\href {\doibase 10.1088/1475-7516/2018/05/052} {\bibfield
  {journal} {\bibinfo  {journal} {JCAP}\ }\textbf {\bibinfo {volume} {05}},\
  \bibinfo {pages} {052} (\bibinfo {year} {2018})},\ \Eprint
  {http://arxiv.org/abs/1802.02281} {arXiv:1802.02281 [gr-qc]} \BibitemShut
  {NoStop}%
\bibitem [{\citenamefont {D'Agostino}\ and\ \citenamefont
  {Nunes}(2023)}]{DAgostino:2023cgx}%
  \BibitemOpen
  \bibfield  {author} {\bibinfo {author} {\bibfnamefont {R.}~\bibnamefont
  {D'Agostino}}\ and\ \bibinfo {author} {\bibfnamefont {R.~C.}\ \bibnamefont
  {Nunes}},\ }\href {\doibase 10.1103/PhysRevD.108.023523} {\bibfield
  {journal} {\bibinfo  {journal} {Phys. Rev. D}\ }\textbf {\bibinfo {volume}
  {108}},\ \bibinfo {pages} {023523} (\bibinfo {year} {2023})},\ \Eprint
  {http://arxiv.org/abs/2307.13464} {arXiv:2307.13464 [astro-ph.CO]}
  \BibitemShut {NoStop}%
\bibitem [{\citenamefont {Braglia}\ \emph {et~al.}(2021)\citenamefont
  {Braglia}, \citenamefont {Ballardini}, \citenamefont {Finelli},\ and\
  \citenamefont {Koyama}}]{Braglia:2020auw}%
  \BibitemOpen
  \bibfield  {author} {\bibinfo {author} {\bibfnamefont {M.}~\bibnamefont
  {Braglia}}, \bibinfo {author} {\bibfnamefont {M.}~\bibnamefont {Ballardini}},
  \bibinfo {author} {\bibfnamefont {F.}~\bibnamefont {Finelli}}, \ and\
  \bibinfo {author} {\bibfnamefont {K.}~\bibnamefont {Koyama}},\ }\href
  {\doibase 10.1103/PhysRevD.103.043528} {\bibfield  {journal} {\bibinfo
  {journal} {Phys. Rev. D}\ }\textbf {\bibinfo {volume} {103}},\ \bibinfo
  {pages} {043528} (\bibinfo {year} {2021})},\ \Eprint
  {http://arxiv.org/abs/2011.12934} {arXiv:2011.12934 [astro-ph.CO]}
  \BibitemShut {NoStop}%
\bibitem [{\citenamefont {Bouch\`e}\ \emph {et~al.}(2023)\citenamefont
  {Bouch\`e}, \citenamefont {Capozziello},\ and\ \citenamefont
  {Salzano}}]{Bouche:2023xjw}%
  \BibitemOpen
  \bibfield  {author} {\bibinfo {author} {\bibfnamefont {F.}~\bibnamefont
  {Bouch\`e}}, \bibinfo {author} {\bibfnamefont {S.}~\bibnamefont
  {Capozziello}}, \ and\ \bibinfo {author} {\bibfnamefont {V.}~\bibnamefont
  {Salzano}},\ }\href {\doibase 10.3390/universe9010027} {\bibfield  {journal}
  {\bibinfo  {journal} {Universe}\ }\textbf {\bibinfo {volume} {9}},\ \bibinfo
  {pages} {27} (\bibinfo {year} {2023})},\ \Eprint
  {http://arxiv.org/abs/2301.01503} {arXiv:2301.01503 [astro-ph.CO]}
  \BibitemShut {NoStop}%
\bibitem [{\citenamefont {Escamilla-Rivera}\ and\ \citenamefont
  {Torres~Castillejos}(2023)}]{Escamilla-Rivera:2022mkc}%
  \BibitemOpen
  \bibfield  {author} {\bibinfo {author} {\bibfnamefont {C.}~\bibnamefont
  {Escamilla-Rivera}}\ and\ \bibinfo {author} {\bibfnamefont {R.}~\bibnamefont
  {Torres~Castillejos}},\ }\href {\doibase 10.3390/universe9010014} {\bibfield
  {journal} {\bibinfo  {journal} {Universe}\ }\textbf {\bibinfo {volume} {9}},\
  \bibinfo {pages} {14} (\bibinfo {year} {2023})},\ \Eprint
  {http://arxiv.org/abs/2301.00490} {arXiv:2301.00490 [astro-ph.CO]}
  \BibitemShut {NoStop}%
\bibitem [{\citenamefont {Petronikolou}\ and\ \citenamefont
  {Saridakis}(2023)}]{Petronikolou:2023cwu}%
  \BibitemOpen
  \bibfield  {author} {\bibinfo {author} {\bibfnamefont {M.}~\bibnamefont
  {Petronikolou}}\ and\ \bibinfo {author} {\bibfnamefont {E.~N.}\ \bibnamefont
  {Saridakis}},\ }\href {\doibase 10.3390/universe9090397} {\bibfield
  {journal} {\bibinfo  {journal} {Universe}\ }\textbf {\bibinfo {volume} {9}},\
  \bibinfo {pages} {397} (\bibinfo {year} {2023})},\ \Eprint
  {http://arxiv.org/abs/2308.16044} {arXiv:2308.16044 [gr-qc]} \BibitemShut
  {NoStop}%
\bibitem [{\citenamefont {Banihashemi}\ \emph {et~al.}(2019)\citenamefont
  {Banihashemi}, \citenamefont {Khosravi},\ and\ \citenamefont
  {Shirazi}}]{Banihashemi:2018has}%
  \BibitemOpen
  \bibfield  {author} {\bibinfo {author} {\bibfnamefont {A.}~\bibnamefont
  {Banihashemi}}, \bibinfo {author} {\bibfnamefont {N.}~\bibnamefont
  {Khosravi}}, \ and\ \bibinfo {author} {\bibfnamefont {A.~H.}\ \bibnamefont
  {Shirazi}},\ }\href {\doibase 10.1103/PhysRevD.99.083509} {\bibfield
  {journal} {\bibinfo  {journal} {Phys. Rev. D}\ }\textbf {\bibinfo {volume}
  {99}},\ \bibinfo {pages} {083509} (\bibinfo {year} {2019})},\ \Eprint
  {http://arxiv.org/abs/1810.11007} {arXiv:1810.11007 [astro-ph.CO]}
  \BibitemShut {NoStop}%
\bibitem [{\citenamefont {Ivanov}\ \emph {et~al.}(2020)\citenamefont {Ivanov},
  \citenamefont {Ali-Ha\"\i{}moud},\ and\ \citenamefont
  {Lesgourgues}}]{Ivanov:2020mfr}%
  \BibitemOpen
  \bibfield  {author} {\bibinfo {author} {\bibfnamefont {M.~M.}\ \bibnamefont
  {Ivanov}}, \bibinfo {author} {\bibfnamefont {Y.}~\bibnamefont
  {Ali-Ha\"\i{}moud}}, \ and\ \bibinfo {author} {\bibfnamefont
  {J.}~\bibnamefont {Lesgourgues}},\ }\href {\doibase
  10.1103/PhysRevD.102.063515} {\bibfield  {journal} {\bibinfo  {journal}
  {Phys. Rev. D}\ }\textbf {\bibinfo {volume} {102}},\ \bibinfo {pages}
  {063515} (\bibinfo {year} {2020})},\ \Eprint
  {http://arxiv.org/abs/2005.10656} {arXiv:2005.10656 [astro-ph.CO]}
  \BibitemShut {NoStop}%
\bibitem [{\citenamefont {Keeley}\ and\ \citenamefont
  {Shafieloo}(2023)}]{Keeley:2022ojz}%
  \BibitemOpen
  \bibfield  {author} {\bibinfo {author} {\bibfnamefont {R.~E.}\ \bibnamefont
  {Keeley}}\ and\ \bibinfo {author} {\bibfnamefont {A.}~\bibnamefont
  {Shafieloo}},\ }\href {\doibase 10.1103/PhysRevLett.131.111002} {\bibfield
  {journal} {\bibinfo  {journal} {Phys. Rev. Lett.}\ }\textbf {\bibinfo
  {volume} {131}},\ \bibinfo {pages} {111002} (\bibinfo {year} {2023})},\
  \Eprint {http://arxiv.org/abs/2206.08440} {arXiv:2206.08440 [astro-ph.CO]}
  \BibitemShut {NoStop}%
\bibitem [{\citenamefont {Bose}\ and\ \citenamefont
  {Lombriser}(2021)}]{Bose:2020cjb}%
  \BibitemOpen
  \bibfield  {author} {\bibinfo {author} {\bibfnamefont {B.}~\bibnamefont
  {Bose}}\ and\ \bibinfo {author} {\bibfnamefont {L.}~\bibnamefont
  {Lombriser}},\ }\href {\doibase 10.1103/PhysRevD.103.L081304} {\bibfield
  {journal} {\bibinfo  {journal} {Phys. Rev. D}\ }\textbf {\bibinfo {volume}
  {103}},\ \bibinfo {pages} {L081304} (\bibinfo {year} {2021})},\ \Eprint
  {http://arxiv.org/abs/2006.16149} {arXiv:2006.16149 [astro-ph.CO]}
  \BibitemShut {NoStop}%
\bibitem [{\citenamefont {Huterer}\ and\ \citenamefont
  {Wu}(2023)}]{Huterer:2023ldv}%
  \BibitemOpen
  \bibfield  {author} {\bibinfo {author} {\bibfnamefont {D.}~\bibnamefont
  {Huterer}}\ and\ \bibinfo {author} {\bibfnamefont {H.-Y.}\ \bibnamefont
  {Wu}},\ }\href@noop {} {\  (\bibinfo {year} {2023})},\ \Eprint
  {http://arxiv.org/abs/2309.05749} {arXiv:2309.05749 [astro-ph.CO]}
  \BibitemShut {NoStop}%
\bibitem [{\citenamefont {Vagnozzi}(2023)}]{Vagnozzi:2023nrq}%
  \BibitemOpen
  \bibfield  {author} {\bibinfo {author} {\bibfnamefont {S.}~\bibnamefont
  {Vagnozzi}},\ }\href {\doibase 10.3390/universe9090393} {\bibfield  {journal}
  {\bibinfo  {journal} {Universe}\ }\textbf {\bibinfo {volume} {9}},\ \bibinfo
  {pages} {393} (\bibinfo {year} {2023})},\ \Eprint
  {http://arxiv.org/abs/2308.16628} {arXiv:2308.16628 [astro-ph.CO]}
  \BibitemShut {NoStop}%
\bibitem [{\citenamefont {Yarahmadi}\ and\ \citenamefont
  {Salehi}(2024)}]{yarahmadi2024using}%
  \BibitemOpen
  \bibfield  {author} {\bibinfo {author} {\bibfnamefont {M.}~\bibnamefont
  {Yarahmadi}}\ and\ \bibinfo {author} {\bibfnamefont {A.}~\bibnamefont
  {Salehi}},\ }\href@noop {} {\bibfield  {journal} {\bibinfo  {journal} {The
  European Physical Journal C}\ }\textbf {\bibinfo {volume} {84}},\ \bibinfo
  {pages} {1} (\bibinfo {year} {2024})}\BibitemShut {NoStop}%
\bibitem [{\citenamefont {Pedreira}\ \emph {et~al.}(2024)\citenamefont
  {Pedreira}, \citenamefont {Benetti}, \citenamefont {Ferreira}, \citenamefont
  {Graef},\ and\ \citenamefont {Herold}}]{Pedreira:2023qqt}%
  \BibitemOpen
  \bibfield  {author} {\bibinfo {author} {\bibfnamefont {I.~d. O.~C.}\
  \bibnamefont {Pedreira}}, \bibinfo {author} {\bibfnamefont {M.}~\bibnamefont
  {Benetti}}, \bibinfo {author} {\bibfnamefont {E.~G.~M.}\ \bibnamefont
  {Ferreira}}, \bibinfo {author} {\bibfnamefont {L.~L.}\ \bibnamefont {Graef}},
  \ and\ \bibinfo {author} {\bibfnamefont {L.}~\bibnamefont {Herold}},\ }\href
  {\doibase 10.1103/PhysRevD.109.103525} {\bibfield  {journal} {\bibinfo
  {journal} {Phys. Rev. D}\ }\textbf {\bibinfo {volume} {109}},\ \bibinfo
  {pages} {103525} (\bibinfo {year} {2024})},\ \Eprint
  {http://arxiv.org/abs/2311.04977} {arXiv:2311.04977 [astro-ph.CO]}
  \BibitemShut {NoStop}%
\bibitem [{\citenamefont {Kable}\ \emph {et~al.}(2023)\citenamefont {Kable},
  \citenamefont {Benevento}, \citenamefont {Addison},\ and\ \citenamefont
  {Bennett}}]{Kable:2023bsg}%
  \BibitemOpen
  \bibfield  {author} {\bibinfo {author} {\bibfnamefont {J.~A.}\ \bibnamefont
  {Kable}}, \bibinfo {author} {\bibfnamefont {G.}~\bibnamefont {Benevento}},
  \bibinfo {author} {\bibfnamefont {G.~E.}\ \bibnamefont {Addison}}, \ and\
  \bibinfo {author} {\bibfnamefont {C.~L.}\ \bibnamefont {Bennett}},\ }\href
  {\doibase 10.3847/1538-4357/acfed0} {\bibfield  {journal} {\bibinfo
  {journal} {Astrophys. J.}\ }\textbf {\bibinfo {volume} {959}},\ \bibinfo
  {pages} {143} (\bibinfo {year} {2023})},\ \Eprint
  {http://arxiv.org/abs/2307.12174} {arXiv:2307.12174 [astro-ph.CO]}
  \BibitemShut {NoStop}%
\bibitem [{\citenamefont {Frion}\ \emph {et~al.}(2023)\citenamefont {Frion},
  \citenamefont {Camarena}, \citenamefont {Giani}, \citenamefont {Miranda},
  \citenamefont {Bertacca}, \citenamefont {Marra},\ and\ \citenamefont
  {Piattella}}]{Frion:2023xwq}%
  \BibitemOpen
  \bibfield  {author} {\bibinfo {author} {\bibfnamefont {E.}~\bibnamefont
  {Frion}}, \bibinfo {author} {\bibfnamefont {D.}~\bibnamefont {Camarena}},
  \bibinfo {author} {\bibfnamefont {L.}~\bibnamefont {Giani}}, \bibinfo
  {author} {\bibfnamefont {T.}~\bibnamefont {Miranda}}, \bibinfo {author}
  {\bibfnamefont {D.}~\bibnamefont {Bertacca}}, \bibinfo {author}
  {\bibfnamefont {V.}~\bibnamefont {Marra}}, \ and\ \bibinfo {author}
  {\bibfnamefont {O.~F.}\ \bibnamefont {Piattella}},\ }\href {\doibase
  10.21105/astro.2307.06320} {\  (\bibinfo {year} {2023}),\
  10.21105/astro.2307.06320},\ \Eprint {http://arxiv.org/abs/2307.06320}
  {arXiv:2307.06320 [astro-ph.CO]} \BibitemShut {NoStop}%
\bibitem [{\citenamefont {Tiwari}\ \emph {et~al.}(2024)\citenamefont {Tiwari},
  \citenamefont {Ghosh},\ and\ \citenamefont {Jain}}]{Tiwari:2023jle}%
  \BibitemOpen
  \bibfield  {author} {\bibinfo {author} {\bibfnamefont {Y.}~\bibnamefont
  {Tiwari}}, \bibinfo {author} {\bibfnamefont {B.}~\bibnamefont {Ghosh}}, \
  and\ \bibinfo {author} {\bibfnamefont {R.~K.}\ \bibnamefont {Jain}},\ }\href
  {\doibase 10.1140/epjc/s10052-024-12577-0} {\bibfield  {journal} {\bibinfo
  {journal} {Eur. Phys. J. C}\ }\textbf {\bibinfo {volume} {84}},\ \bibinfo
  {pages} {220} (\bibinfo {year} {2024})},\ \Eprint
  {http://arxiv.org/abs/2301.09382} {arXiv:2301.09382 [astro-ph.CO]}
  \BibitemShut {NoStop}%
\bibitem [{\citenamefont {Jedamzik}\ and\ \citenamefont
  {Pogosian}(2020)}]{Jedamzik:2020krr}%
  \BibitemOpen
  \bibfield  {author} {\bibinfo {author} {\bibfnamefont {K.}~\bibnamefont
  {Jedamzik}}\ and\ \bibinfo {author} {\bibfnamefont {L.}~\bibnamefont
  {Pogosian}},\ }\href {\doibase 10.1103/PhysRevLett.125.181302} {\bibfield
  {journal} {\bibinfo  {journal} {Phys. Rev. Lett.}\ }\textbf {\bibinfo
  {volume} {125}},\ \bibinfo {pages} {181302} (\bibinfo {year} {2020})},\
  \Eprint {http://arxiv.org/abs/2004.09487} {arXiv:2004.09487 [astro-ph.CO]}
  \BibitemShut {NoStop}%
\bibitem [{\citenamefont {Dainotti}\ \emph {et~al.}(2021)\citenamefont
  {Dainotti}, \citenamefont {De~Simone}, \citenamefont {Schiavone},
  \citenamefont {Montani}, \citenamefont {Rinaldi},\ and\ \citenamefont
  {Lambiase}}]{Dainotti:2021pqg}%
  \BibitemOpen
  \bibfield  {author} {\bibinfo {author} {\bibfnamefont {M.~G.}\ \bibnamefont
  {Dainotti}}, \bibinfo {author} {\bibfnamefont {B.}~\bibnamefont {De~Simone}},
  \bibinfo {author} {\bibfnamefont {T.}~\bibnamefont {Schiavone}}, \bibinfo
  {author} {\bibfnamefont {G.}~\bibnamefont {Montani}}, \bibinfo {author}
  {\bibfnamefont {E.}~\bibnamefont {Rinaldi}}, \ and\ \bibinfo {author}
  {\bibfnamefont {G.}~\bibnamefont {Lambiase}},\ }\href {\doibase
  10.3847/1538-4357/abeb73} {\bibfield  {journal} {\bibinfo  {journal}
  {Astrophys. J.}\ }\textbf {\bibinfo {volume} {912}},\ \bibinfo {pages} {150}
  (\bibinfo {year} {2021})},\ \Eprint {http://arxiv.org/abs/2103.02117}
  {arXiv:2103.02117 [astro-ph.CO]} \BibitemShut {NoStop}%
\bibitem [{\citenamefont {Dainotti}\ \emph {et~al.}(2022)\citenamefont
  {Dainotti}, \citenamefont {De~Simone}, \citenamefont {Schiavone},
  \citenamefont {Montani}, \citenamefont {Rinaldi}, \citenamefont {Lambiase},
  \citenamefont {Bogdan},\ and\ \citenamefont {Ugale}}]{Dainotti:2022bzg}%
  \BibitemOpen
  \bibfield  {author} {\bibinfo {author} {\bibfnamefont {M.~G.}\ \bibnamefont
  {Dainotti}}, \bibinfo {author} {\bibfnamefont {B.}~\bibnamefont {De~Simone}},
  \bibinfo {author} {\bibfnamefont {T.}~\bibnamefont {Schiavone}}, \bibinfo
  {author} {\bibfnamefont {G.}~\bibnamefont {Montani}}, \bibinfo {author}
  {\bibfnamefont {E.}~\bibnamefont {Rinaldi}}, \bibinfo {author} {\bibfnamefont
  {G.}~\bibnamefont {Lambiase}}, \bibinfo {author} {\bibfnamefont
  {M.}~\bibnamefont {Bogdan}}, \ and\ \bibinfo {author} {\bibfnamefont
  {S.}~\bibnamefont {Ugale}},\ }\href {\doibase 10.3390/galaxies10010024}
  {\bibfield  {journal} {\bibinfo  {journal} {Galaxies}\ }\textbf {\bibinfo
  {volume} {10}},\ \bibinfo {pages} {24} (\bibinfo {year} {2022})},\ \Eprint
  {http://arxiv.org/abs/2201.09848} {arXiv:2201.09848 [astro-ph.CO]}
  \BibitemShut {NoStop}%
\bibitem [{\citenamefont {Ade}\ \emph {et~al.}(2021{\natexlab{a}})\citenamefont
  {Ade} \emph {et~al.}}]{BICEP:2021xfz}%
  \BibitemOpen
  \bibfield  {author} {\bibinfo {author} {\bibfnamefont {P.~A.~R.}\
  \bibnamefont {Ade}} \emph {et~al.} (\bibinfo {collaboration} {BICEP, Keck}),\
  }\href {\doibase 10.1103/PhysRevLett.127.151301} {\bibfield  {journal}
  {\bibinfo  {journal} {Phys. Rev. Lett.}\ }\textbf {\bibinfo {volume} {127}},\
  \bibinfo {pages} {151301} (\bibinfo {year} {2021}{\natexlab{a}})},\ \Eprint
  {http://arxiv.org/abs/2110.00483} {arXiv:2110.00483 [astro-ph.CO]}
  \BibitemShut {NoStop}%
\bibitem [{\citenamefont {Linde}(1983)}]{linde1983chaotic}%
  \BibitemOpen
  \bibfield  {author} {\bibinfo {author} {\bibfnamefont {A.~D.}\ \bibnamefont
  {Linde}},\ }\href@noop {} {\bibfield  {journal} {\bibinfo  {journal} {Physics
  Letters B}\ }\textbf {\bibinfo {volume} {129}},\ \bibinfo {pages} {177}
  (\bibinfo {year} {1983})}\BibitemShut {NoStop}%
\bibitem [{\citenamefont {Armendariz-Picon}\ \emph {et~al.}(1999)\citenamefont
  {Armendariz-Picon}, \citenamefont {Damour},\ and\ \citenamefont
  {Mukhanov}}]{Armendariz-Picon:1999hyi}%
  \BibitemOpen
  \bibfield  {author} {\bibinfo {author} {\bibfnamefont {C.}~\bibnamefont
  {Armendariz-Picon}}, \bibinfo {author} {\bibfnamefont {T.}~\bibnamefont
  {Damour}}, \ and\ \bibinfo {author} {\bibfnamefont {V.~F.}\ \bibnamefont
  {Mukhanov}},\ }\href {\doibase 10.1016/S0370-2693(99)00603-6} {\bibfield
  {journal} {\bibinfo  {journal} {Phys. Lett. B}\ }\textbf {\bibinfo {volume}
  {458}},\ \bibinfo {pages} {209} (\bibinfo {year} {1999})},\ \Eprint
  {http://arxiv.org/abs/hep-th/9904075} {arXiv:hep-th/9904075} \BibitemShut
  {NoStop}%
\bibitem [{\citenamefont {Armendariz-Picon}\ \emph {et~al.}(2001)\citenamefont
  {Armendariz-Picon}, \citenamefont {Mukhanov},\ and\ \citenamefont
  {Steinhardt}}]{Armendariz-Picon:2000ulo}%
  \BibitemOpen
  \bibfield  {author} {\bibinfo {author} {\bibfnamefont {C.}~\bibnamefont
  {Armendariz-Picon}}, \bibinfo {author} {\bibfnamefont {V.~F.}\ \bibnamefont
  {Mukhanov}}, \ and\ \bibinfo {author} {\bibfnamefont {P.~J.}\ \bibnamefont
  {Steinhardt}},\ }\href {\doibase 10.1103/PhysRevD.63.103510} {\bibfield
  {journal} {\bibinfo  {journal} {Phys. Rev. D}\ }\textbf {\bibinfo {volume}
  {63}},\ \bibinfo {pages} {103510} (\bibinfo {year} {2001})},\ \Eprint
  {http://arxiv.org/abs/astro-ph/0006373} {arXiv:astro-ph/0006373} \BibitemShut
  {NoStop}%
\bibitem [{\citenamefont {Hosseini~Mansoori}\ \emph
  {et~al.}(2023{\natexlab{a}})\citenamefont {Hosseini~Mansoori}, \citenamefont
  {Felegary}, \citenamefont {Roshan}, \citenamefont {Akarsu},\ and\
  \citenamefont {Sami}}]{HosseiniMansoori:2023zop}%
  \BibitemOpen
  \bibfield  {author} {\bibinfo {author} {\bibfnamefont {S.~A.}\ \bibnamefont
  {Hosseini~Mansoori}}, \bibinfo {author} {\bibfnamefont {F.}~\bibnamefont
  {Felegary}}, \bibinfo {author} {\bibfnamefont {M.}~\bibnamefont {Roshan}},
  \bibinfo {author} {\bibfnamefont {O.}~\bibnamefont {Akarsu}}, \ and\ \bibinfo
  {author} {\bibfnamefont {M.}~\bibnamefont {Sami}},\ }\href {\doibase
  10.1016/j.dark.2023.101360} {\bibfield  {journal} {\bibinfo  {journal} {Phys.
  Dark Univ.}\ }\textbf {\bibinfo {volume} {42}},\ \bibinfo {pages} {101360}
  (\bibinfo {year} {2023}{\natexlab{a}})},\ \Eprint
  {http://arxiv.org/abs/2306.09181} {arXiv:2306.09181 [gr-qc]} \BibitemShut
  {NoStop}%
\bibitem [{\citenamefont {Hosseini~Mansoori}\ \emph
  {et~al.}(2023{\natexlab{b}})\citenamefont {Hosseini~Mansoori}, \citenamefont
  {Felegray}, \citenamefont {Talebian},\ and\ \citenamefont
  {Sami}}]{HosseiniMansoori:2023mqh}%
  \BibitemOpen
  \bibfield  {author} {\bibinfo {author} {\bibfnamefont {S.~A.}\ \bibnamefont
  {Hosseini~Mansoori}}, \bibinfo {author} {\bibfnamefont {F.}~\bibnamefont
  {Felegray}}, \bibinfo {author} {\bibfnamefont {A.}~\bibnamefont {Talebian}},
  \ and\ \bibinfo {author} {\bibfnamefont {M.}~\bibnamefont {Sami}},\ }\href
  {\doibase 10.1088/1475-7516/2023/08/067} {\bibfield  {journal} {\bibinfo
  {journal} {JCAP}\ }\textbf {\bibinfo {volume} {08}},\ \bibinfo {pages} {067}
  (\bibinfo {year} {2023}{\natexlab{b}})},\ \Eprint
  {http://arxiv.org/abs/2307.06757} {arXiv:2307.06757 [astro-ph.CO]}
  \BibitemShut {NoStop}%
\bibitem [{\citenamefont {Chen}\ \emph
  {et~al.}(2007{\natexlab{a}})\citenamefont {Chen}, \citenamefont {Huang},
  \citenamefont {Kachru},\ and\ \citenamefont {Shiu}}]{Chen:2006nt}%
  \BibitemOpen
  \bibfield  {author} {\bibinfo {author} {\bibfnamefont {X.}~\bibnamefont
  {Chen}}, \bibinfo {author} {\bibfnamefont {M.-x.}\ \bibnamefont {Huang}},
  \bibinfo {author} {\bibfnamefont {S.}~\bibnamefont {Kachru}}, \ and\ \bibinfo
  {author} {\bibfnamefont {G.}~\bibnamefont {Shiu}},\ }\href {\doibase
  10.1088/1475-7516/2007/01/002} {\bibfield  {journal} {\bibinfo  {journal}
  {JCAP}\ }\textbf {\bibinfo {volume} {01}},\ \bibinfo {pages} {002} (\bibinfo
  {year} {2007}{\natexlab{a}})},\ \Eprint {http://arxiv.org/abs/hep-th/0605045}
  {arXiv:hep-th/0605045} \BibitemShut {NoStop}%
\bibitem [{\citenamefont {Maldacena}(2003)}]{maldacena2003non}%
  \BibitemOpen
  \bibfield  {author} {\bibinfo {author} {\bibfnamefont {J.~M.}\ \bibnamefont
  {Maldacena}},\ }\href {\doibase 10.1088/1126-6708/2003/05/013} {\bibfield
  {journal} {\bibinfo  {journal} {JHEP}\ }\textbf {\bibinfo {volume} {05}},\
  \bibinfo {pages} {013} (\bibinfo {year} {2003})},\ \Eprint
  {http://arxiv.org/abs/astro-ph/0210603} {arXiv:astro-ph/0210603} \BibitemShut
  {NoStop}%
\bibitem [{\citenamefont {Chen}\ \emph
  {et~al.}(2007{\natexlab{b}})\citenamefont {Chen}, \citenamefont {Huang},
  \citenamefont {Kachru},\ and\ \citenamefont {Shiu}}]{chen2007observational}%
  \BibitemOpen
  \bibfield  {author} {\bibinfo {author} {\bibfnamefont {X.}~\bibnamefont
  {Chen}}, \bibinfo {author} {\bibfnamefont {M.-x.}\ \bibnamefont {Huang}},
  \bibinfo {author} {\bibfnamefont {S.}~\bibnamefont {Kachru}}, \ and\ \bibinfo
  {author} {\bibfnamefont {G.}~\bibnamefont {Shiu}},\ }\href {\doibase
  10.1088/1475-7516/2007/01/002} {\bibfield  {journal} {\bibinfo  {journal}
  {JCAP}\ }\textbf {\bibinfo {volume} {01}},\ \bibinfo {pages} {002} (\bibinfo
  {year} {2007}{\natexlab{b}})},\ \Eprint {http://arxiv.org/abs/hep-th/0605045}
  {arXiv:hep-th/0605045} \BibitemShut {NoStop}%
\bibitem [{\citenamefont {Seery}\ and\ \citenamefont
  {Lidsey}(2005)}]{Seery:2005wm}%
  \BibitemOpen
  \bibfield  {author} {\bibinfo {author} {\bibfnamefont {D.}~\bibnamefont
  {Seery}}\ and\ \bibinfo {author} {\bibfnamefont {J.~E.}\ \bibnamefont
  {Lidsey}},\ }\href {\doibase 10.1088/1475-7516/2005/06/003} {\bibfield
  {journal} {\bibinfo  {journal} {JCAP}\ }\textbf {\bibinfo {volume} {06}},\
  \bibinfo {pages} {003} (\bibinfo {year} {2005})},\ \Eprint
  {http://arxiv.org/abs/astro-ph/0503692} {arXiv:astro-ph/0503692} \BibitemShut
  {NoStop}%
\bibitem [{\citenamefont {Li}\ and\ \citenamefont {Liddle}(2012)}]{Li:2012vta}%
  \BibitemOpen
  \bibfield  {author} {\bibinfo {author} {\bibfnamefont {S.}~\bibnamefont
  {Li}}\ and\ \bibinfo {author} {\bibfnamefont {A.~R.}\ \bibnamefont
  {Liddle}},\ }\href {\doibase 10.1088/1475-7516/2012/10/011} {\bibfield
  {journal} {\bibinfo  {journal} {JCAP}\ }\textbf {\bibinfo {volume} {10}},\
  \bibinfo {pages} {011} (\bibinfo {year} {2012})},\ \Eprint
  {http://arxiv.org/abs/1204.6214} {arXiv:1204.6214 [astro-ph.CO]} \BibitemShut
  {NoStop}%
\bibitem [{\citenamefont {Unnikrishnan}\ \emph {et~al.}(2012)\citenamefont
  {Unnikrishnan}, \citenamefont {Sahni},\ and\ \citenamefont
  {Toporensky}}]{Unnikrishnan:2012zu}%
  \BibitemOpen
  \bibfield  {author} {\bibinfo {author} {\bibfnamefont {S.}~\bibnamefont
  {Unnikrishnan}}, \bibinfo {author} {\bibfnamefont {V.}~\bibnamefont {Sahni}},
  \ and\ \bibinfo {author} {\bibfnamefont {A.}~\bibnamefont {Toporensky}},\
  }\href {\doibase 10.1088/1475-7516/2012/08/018} {\bibfield  {journal}
  {\bibinfo  {journal} {JCAP}\ }\textbf {\bibinfo {volume} {08}},\ \bibinfo
  {pages} {018} (\bibinfo {year} {2012})},\ \Eprint
  {http://arxiv.org/abs/1205.0786} {arXiv:1205.0786 [astro-ph.CO]} \BibitemShut
  {NoStop}%
\bibitem [{\citenamefont {Akrami}\ \emph {et~al.}(2020)\citenamefont {Akrami}
  \emph {et~al.}}]{Planck:2018jri}%
  \BibitemOpen
  \bibfield  {author} {\bibinfo {author} {\bibfnamefont {Y.}~\bibnamefont
  {Akrami}} \emph {et~al.} (\bibinfo {collaboration} {Planck}),\ }\href
  {\doibase 10.1051/0004-6361/201833887} {\bibfield  {journal} {\bibinfo
  {journal} {Astron. Astrophys.}\ }\textbf {\bibinfo {volume} {641}},\ \bibinfo
  {pages} {A10} (\bibinfo {year} {2020})},\ \Eprint
  {http://arxiv.org/abs/1807.06211} {arXiv:1807.06211 [astro-ph.CO]}
  \BibitemShut {NoStop}%
\bibitem [{\citenamefont {Ade}\ \emph {et~al.}(2018)\citenamefont {Ade} \emph
  {et~al.}}]{BICEP2:2018kqh}%
  \BibitemOpen
  \bibfield  {author} {\bibinfo {author} {\bibfnamefont {P.~A.~R.}\
  \bibnamefont {Ade}} \emph {et~al.} (\bibinfo {collaboration} {BICEP2, Keck
  Array}),\ }\href {\doibase 10.1103/PhysRevLett.121.221301} {\bibfield
  {journal} {\bibinfo  {journal} {Phys. Rev. Lett.}\ }\textbf {\bibinfo
  {volume} {121}},\ \bibinfo {pages} {221301} (\bibinfo {year} {2018})},\
  \Eprint {http://arxiv.org/abs/1810.05216} {arXiv:1810.05216 [astro-ph.CO]}
  \BibitemShut {NoStop}%
\bibitem [{\citenamefont {Aghanim}\ \emph
  {et~al.}(2020{\natexlab{b}})\citenamefont {Aghanim} \emph
  {et~al.}}]{aghanim2020planck}%
  \BibitemOpen
  \bibfield  {author} {\bibinfo {author} {\bibfnamefont {N.}~\bibnamefont
  {Aghanim}} \emph {et~al.} (\bibinfo {collaboration} {Planck}),\ }\href
  {\doibase 10.1051/0004-6361/201833910} {\bibfield  {journal} {\bibinfo
  {journal} {Astron. Astrophys.}\ }\textbf {\bibinfo {volume} {641}},\ \bibinfo
  {pages} {A6} (\bibinfo {year} {2020}{\natexlab{b}})},\ \bibinfo {note}
  {[Erratum: Astron.Astrophys. 652, C4 (2021)]},\ \Eprint
  {http://arxiv.org/abs/1807.06209} {arXiv:1807.06209 [astro-ph.CO]}
  \BibitemShut {NoStop}%
\bibitem [{\citenamefont {Ade}\ \emph {et~al.}(2021{\natexlab{b}})\citenamefont
  {Ade} \emph {et~al.}}]{ade2021improved}%
  \BibitemOpen
  \bibfield  {author} {\bibinfo {author} {\bibfnamefont {P.~A.~R.}\
  \bibnamefont {Ade}} \emph {et~al.} (\bibinfo {collaboration} {BICEP, Keck}),\
  }\href {\doibase 10.1103/PhysRevLett.127.151301} {\bibfield  {journal}
  {\bibinfo  {journal} {Phys. Rev. Lett.}\ }\textbf {\bibinfo {volume} {127}},\
  \bibinfo {pages} {151301} (\bibinfo {year} {2021}{\natexlab{b}})},\ \Eprint
  {http://arxiv.org/abs/2110.00483} {arXiv:2110.00483 [astro-ph.CO]}
  \BibitemShut {NoStop}%
\bibitem [{\citenamefont {Lewis}\ \emph {et~al.}(2000)\citenamefont {Lewis},
  \citenamefont {Challinor},\ and\ \citenamefont {Lasenby}}]{Lewis:1999bs}%
  \BibitemOpen
  \bibfield  {author} {\bibinfo {author} {\bibfnamefont {A.}~\bibnamefont
  {Lewis}}, \bibinfo {author} {\bibfnamefont {A.}~\bibnamefont {Challinor}}, \
  and\ \bibinfo {author} {\bibfnamefont {A.}~\bibnamefont {Lasenby}},\ }\href
  {\doibase 10.1086/309179} {\bibfield  {journal} {\bibinfo  {journal}
  {Astrophys. J.}\ }\textbf {\bibinfo {volume} {538}},\ \bibinfo {pages} {473}
  (\bibinfo {year} {2000})},\ \Eprint {http://arxiv.org/abs/astro-ph/9911177}
  {arXiv:astro-ph/9911177} \BibitemShut {NoStop}%
\bibitem [{\citenamefont {{Howlett}}\ \emph {et~al.}(2012)\citenamefont
  {{Howlett}}, \citenamefont {{Lewis}}, \citenamefont {{Hall}},\ and\
  \citenamefont {{Challinor}}}]{Howlett2012}%
  \BibitemOpen
  \bibfield  {author} {\bibinfo {author} {\bibfnamefont {C.}~\bibnamefont
  {{Howlett}}}, \bibinfo {author} {\bibfnamefont {A.}~\bibnamefont {{Lewis}}},
  \bibinfo {author} {\bibfnamefont {A.}~\bibnamefont {{Hall}}}, \ and\ \bibinfo
  {author} {\bibfnamefont {A.}~\bibnamefont {{Challinor}}},\ }\href {\doibase
  10.1088/1475-7516/2012/04/027} {\bibfield  {journal} {\bibinfo  {journal}
  {JCAP}\ }\textbf {\bibinfo {volume} {2012}},\ \bibinfo {eid} {027} (\bibinfo
  {year} {2012})},\ \Eprint {http://arxiv.org/abs/1201.3654} {arXiv:1201.3654
  [astro-ph.CO]} \BibitemShut {NoStop}%
\bibitem [{\citenamefont {Lewis}\ and\ \citenamefont
  {Bridle}(2002)}]{Lewis:2002ah}%
  \BibitemOpen
  \bibfield  {author} {\bibinfo {author} {\bibfnamefont {A.}~\bibnamefont
  {Lewis}}\ and\ \bibinfo {author} {\bibfnamefont {S.}~\bibnamefont {Bridle}},\
  }\href {\doibase 10.1103/PhysRevD.66.103511} {\bibfield  {journal} {\bibinfo
  {journal} {Phys. Rev. D}\ }\textbf {\bibinfo {volume} {66}},\ \bibinfo
  {pages} {103511} (\bibinfo {year} {2002})},\ \Eprint
  {http://arxiv.org/abs/astro-ph/0205436} {arXiv:astro-ph/0205436} \BibitemShut
  {NoStop}%
\bibitem [{\citenamefont {Aghanim}\ \emph
  {et~al.}(2020{\natexlab{c}})\citenamefont {Aghanim} \emph
  {et~al.}}]{Planck:2018lbu}%
  \BibitemOpen
  \bibfield  {author} {\bibinfo {author} {\bibfnamefont {N.}~\bibnamefont
  {Aghanim}} \emph {et~al.} (\bibinfo {collaboration} {Planck}),\ }\href
  {\doibase 10.1051/0004-6361/201833886} {\bibfield  {journal} {\bibinfo
  {journal} {Astron. Astrophys.}\ }\textbf {\bibinfo {volume} {641}},\ \bibinfo
  {pages} {A8} (\bibinfo {year} {2020}{\natexlab{c}})},\ \Eprint
  {http://arxiv.org/abs/1807.06210} {arXiv:1807.06210 [astro-ph.CO]}
  \BibitemShut {NoStop}%
\bibitem [{\citenamefont {Aghanim}\ \emph
  {et~al.}(2020{\natexlab{d}})\citenamefont {Aghanim} \emph
  {et~al.}}]{Planck:2019nip}%
  \BibitemOpen
  \bibfield  {author} {\bibinfo {author} {\bibfnamefont {N.}~\bibnamefont
  {Aghanim}} \emph {et~al.} (\bibinfo {collaboration} {Planck}),\ }\href
  {\doibase 10.1051/0004-6361/201936386} {\bibfield  {journal} {\bibinfo
  {journal} {Astron. Astrophys.}\ }\textbf {\bibinfo {volume} {641}},\ \bibinfo
  {pages} {A5} (\bibinfo {year} {2020}{\natexlab{d}})},\ \Eprint
  {http://arxiv.org/abs/1907.12875} {arXiv:1907.12875 [astro-ph.CO]}
  \BibitemShut {NoStop}%
\bibitem [{\citenamefont {Beutler}\ \emph {et~al.}(2011)\citenamefont
  {Beutler}, \citenamefont {Blake}, \citenamefont {Colless}, \citenamefont
  {Jones}, \citenamefont {Staveley-Smith}, \citenamefont {Campbell},
  \citenamefont {Parker}, \citenamefont {Saunders},\ and\ \citenamefont
  {Watson}}]{Beutler:2011hx}%
  \BibitemOpen
  \bibfield  {author} {\bibinfo {author} {\bibfnamefont {F.}~\bibnamefont
  {Beutler}}, \bibinfo {author} {\bibfnamefont {C.}~\bibnamefont {Blake}},
  \bibinfo {author} {\bibfnamefont {M.}~\bibnamefont {Colless}}, \bibinfo
  {author} {\bibfnamefont {D.~H.}\ \bibnamefont {Jones}}, \bibinfo {author}
  {\bibfnamefont {L.}~\bibnamefont {Staveley-Smith}}, \bibinfo {author}
  {\bibfnamefont {L.}~\bibnamefont {Campbell}}, \bibinfo {author}
  {\bibfnamefont {Q.}~\bibnamefont {Parker}}, \bibinfo {author} {\bibfnamefont
  {W.}~\bibnamefont {Saunders}}, \ and\ \bibinfo {author} {\bibfnamefont
  {F.}~\bibnamefont {Watson}},\ }\href {\doibase
  10.1111/j.1365-2966.2011.19250.x} {\bibfield  {journal} {\bibinfo  {journal}
  {Mon. Not. Roy. Astron. Soc.}\ }\textbf {\bibinfo {volume} {416}},\ \bibinfo
  {pages} {3017} (\bibinfo {year} {2011})},\ \Eprint
  {http://arxiv.org/abs/1106.3366} {arXiv:1106.3366 [astro-ph.CO]} \BibitemShut
  {NoStop}%
\bibitem [{\citenamefont {Ross}\ \emph {et~al.}(2015)\citenamefont {Ross},
  \citenamefont {Samushia}, \citenamefont {Howlett}, \citenamefont {Percival},
  \citenamefont {Burden},\ and\ \citenamefont {Manera}}]{Ross:2014qpa}%
  \BibitemOpen
  \bibfield  {author} {\bibinfo {author} {\bibfnamefont {A.~J.}\ \bibnamefont
  {Ross}}, \bibinfo {author} {\bibfnamefont {L.}~\bibnamefont {Samushia}},
  \bibinfo {author} {\bibfnamefont {C.}~\bibnamefont {Howlett}}, \bibinfo
  {author} {\bibfnamefont {W.~J.}\ \bibnamefont {Percival}}, \bibinfo {author}
  {\bibfnamefont {A.}~\bibnamefont {Burden}}, \ and\ \bibinfo {author}
  {\bibfnamefont {M.}~\bibnamefont {Manera}},\ }\href {\doibase
  10.1093/mnras/stv154} {\bibfield  {journal} {\bibinfo  {journal} {Mon. Not.
  Roy. Astron. Soc.}\ }\textbf {\bibinfo {volume} {449}},\ \bibinfo {pages}
  {835} (\bibinfo {year} {2015})},\ \Eprint {http://arxiv.org/abs/1409.3242}
  {arXiv:1409.3242 [astro-ph.CO]} \BibitemShut {NoStop}%
\bibitem [{\citenamefont {Alam}\ \emph {et~al.}(2017)\citenamefont {Alam} \emph
  {et~al.}}]{BOSS:2016wmc}%
  \BibitemOpen
  \bibfield  {author} {\bibinfo {author} {\bibfnamefont {S.}~\bibnamefont
  {Alam}} \emph {et~al.} (\bibinfo {collaboration} {BOSS}),\ }\href {\doibase
  10.1093/mnras/stx721} {\bibfield  {journal} {\bibinfo  {journal} {Mon. Not.
  Roy. Astron. Soc.}\ }\textbf {\bibinfo {volume} {470}},\ \bibinfo {pages}
  {2617} (\bibinfo {year} {2017})},\ \Eprint {http://arxiv.org/abs/1607.03155}
  {arXiv:1607.03155 [astro-ph.CO]} \BibitemShut {NoStop}%
\bibitem [{\citenamefont {Scolnic}\ \emph {et~al.}(2018)\citenamefont {Scolnic}
  \emph {et~al.}}]{Pan-STARRS1:2017jku}%
  \BibitemOpen
  \bibfield  {author} {\bibinfo {author} {\bibfnamefont {D.~M.}\ \bibnamefont
  {Scolnic}} \emph {et~al.} (\bibinfo {collaboration} {Pan-STARRS1}),\ }\href
  {\doibase 10.3847/1538-4357/aab9bb} {\bibfield  {journal} {\bibinfo
  {journal} {Astrophys. J.}\ }\textbf {\bibinfo {volume} {859}},\ \bibinfo
  {pages} {101} (\bibinfo {year} {2018})},\ \Eprint
  {http://arxiv.org/abs/1710.00845} {arXiv:1710.00845 [astro-ph.CO]}
  \BibitemShut {NoStop}%
\bibitem [{\citenamefont {Raveri}\ and\ \citenamefont
  {Hu}(2019)}]{Raveri:2018wln}%
  \BibitemOpen
  \bibfield  {author} {\bibinfo {author} {\bibfnamefont {M.}~\bibnamefont
  {Raveri}}\ and\ \bibinfo {author} {\bibfnamefont {W.}~\bibnamefont {Hu}},\
  }\href {\doibase 10.1103/PhysRevD.99.043506} {\bibfield  {journal} {\bibinfo
  {journal} {Phys. Rev. D}\ }\textbf {\bibinfo {volume} {99}},\ \bibinfo
  {pages} {043506} (\bibinfo {year} {2019})},\ \Eprint
  {http://arxiv.org/abs/1806.04649} {arXiv:1806.04649 [astro-ph.CO]}
  \BibitemShut {NoStop}%
\bibitem [{\citenamefont {Moshafi}\ \emph {et~al.}(2024)\citenamefont
  {Moshafi}, \citenamefont {Talebian}, \citenamefont {Yusofi},\ and\
  \citenamefont {Di~Valentino}}]{Moshafi:2024guo}%
  \BibitemOpen
  \bibfield  {author} {\bibinfo {author} {\bibfnamefont {H.}~\bibnamefont
  {Moshafi}}, \bibinfo {author} {\bibfnamefont {A.}~\bibnamefont {Talebian}},
  \bibinfo {author} {\bibfnamefont {E.}~\bibnamefont {Yusofi}}, \ and\ \bibinfo
  {author} {\bibfnamefont {E.}~\bibnamefont {Di~Valentino}},\ }\href@noop {}
  {\bibfield  {journal} {\bibinfo  {journal} {Phys. Dark Univ.}\ } (\bibinfo
  {year} {2024})},\ \Eprint {http://arxiv.org/abs/2403.02000} {arXiv:2403.02000
  [astro-ph.CO]} \BibitemShut {NoStop}%
\end{thebibliography}%
	
\end{document}